%% file: index.tex
\documentclass[letterpaper,12pt]{article}
\usepackage[left=2.5cm,right=2.5cm,top=2.75cm,bottom=2.5cm,nohead]{geometry}

\def\myemph#1{{\emph{#1}}}
\def\etal{{\myemph{et al}}}

\date{}

\title{On the Technology Prospects and Investment Opportunities for Scalable Neuroscience}

\author{Thomas Dean$^{\mbox{\bf{1,2,3}}}$ \and Biafra Ahanonu$^{\mbox{\bf{3}}}$ \and Mainak Chowdhury$^{\mbox{\bf{3}}}$ \and Anjali Datta$^{\mbox{\bf{3}}}$ \and Andre Esteva$^{\mbox{\bf{3}}}$ \and Daniel Eth$^{\mbox{\bf{3}}}$ \and Nobie Redmon$^{\mbox{\bf{3}}}$ \and Oleg Rumyantsev$^{\mbox{\bf{3}}}$ \hspace{0.25in} Ysis Tarter$^{\mbox{\bf{3}}}$\\\vspace{0.5in}\\{\bf{1}} Google Research, {\bf{2}} Brown University, {\bf{3}} Stanford University}

\begin{document}

% Title and Authors:
\maketitle
\thispagestyle{empty}

% PDF VERSION
\newpage
	
% Table of Contents:
\pagenumbering{roman}
\tableofcontents

% PDF VERSION
\newpage
\pagenumbering{arabic}

% Executive Summary: 
\input{executing.tex}
% Section~\ref{section:executing}

% PDF VERSION
\newpage

% Technical Preliminaries:
\input{introducing.tex}

% Section~\ref{section:introducing}

% Imaging Technologies:
\input{imaging.tex}
% Section~\ref{section:imaging}

% Recording Technologies:
\input{probing.tex}
% Section~\ref{section:probing}

% Automating Neuroscience:
\input{automating.tex}
% Section~\ref{section:automating}

% Synthetic Neurobiology: 
\input{sequencing.tex}
% Section~\ref{section:sequencing}

% Nanotechnology:
\input{shrinking.tex}

% Section~\ref{section:shrinking}

% Investment Opportunities:
\input{investing.tex}

% Section~\ref{section:investing}

% Acknowledgements:
\input{acknowledging.tex}

% Section~\ref{section:acknowledging}

\appendix
\section{Leveraging Sequencing for Recording --- Biafra Ahanonu}
\label{appendix:biafra}
\input{biafra.tex}
\section{Scalable Analytics and Data Mining --- Mainak Chowdhury}
\label{appendix:mainak}
\input{mainak.tex}
\section{Macroscale Imaging Technologies --- Anjali Datta}
\label{appendix:anjali}
\input{anjali.tex}
\section{Nanoscale Recording and Wireless Readout --- Andre Esteva}
\label{appendix:andre}
\input{andre.tex}
\section{Hybrid Biological and Nanotechnology Solutions --- Daniel Eth}
\label{appendix:daniel}
\input{daniel.tex}
\section{Advances in Contrast Agents and Tissue Preparation --- Nobie Redmon}
\label{appendix:nobie}
\input{nobie.tex}
\section{Microendoscopy and Optically Coupled Implants --- Oleg Rumyantsev}
\label{appendix:oleg}
\input{oleg.tex}
\section{Opportunities for Automating Laboratory Procedures   --- Ysis Tarter}
\label{appendix:ysis}
\input{ysis.tex}

\end{document}

%% file: executing.tex
\section{Executive Summary}
\label{section:executing}

% This section was adapted from my talk in the Fujitsu Technology Symposium at the Computer History Museum on June 5, 2013.

Two major initiatives to accelerate research in the brain sciences have focused attention on developing a new generation of scientific instruments for neuroscience.  These instruments will be used to record static (structural) and dynamic (behavioral) information at unprecedented spatial and temporal resolution and report out that information in a form suitable for computational analysis. We distinguish between recording --- taking measurements of individual cells and the extracellular matrix --- and reporting --- transcoding, packaging and transmitting the resulting information for subsequent analysis --- as these represent very different challenges as we scale the relevant technologies to support simultaneously tracking the many neurons that comprise neural circuits of interest. We investigate a diverse set of technologies with the purpose of anticipating their development over the span of the next 10 years and categorizing their impact in terms of short-term [1-2 years], medium-term [2-5 years] and longer-term [5-10 years] deliverables. 
\begin{itemize}
\item short-term options [1--2 years] ---
  The most powerful recording and reporting technologies currently available all use some form of imaging in which some portion of the acoustic or electromagnetic spectrum is used to illuminate a target tissue and the resulting response, attenuated by absorption, reflection and scattering, is analyzed to extract useful information about structure, e.g., cytoarchitectural details, and function, e.g., membrane potentials and spike-timing data. These relatively mature technologies largely finesse the problems relating to powering reporting devices and carrying out the required computations involved in signal processing, compression and transmission by performing all these functions external to the brain. Example technologies include electroencephalography, focused ultrasound, magnetic resonance imaging, microendoscopy, photoacoustic imaging, two-photon calcium imaging, array tomography for proteomics, immunoflorescence for genomics and light-sheet fluorescence microscopy. This class of technologies also includes our current best non-invasive options for the study of human subjects. 

  Incremental improvements in these technologies are likely to continue unabated for some time, enabled by advances in biology and physics and funded and motivated by applications in medicine and materials science.  In order to better resolve features of interest, biochemists are developing new reagents that are differentially absorbed by cellular structures and serve to alter the local spectral characteristics of illuminated tissue.  Tissue samples can be prepared in such a way that structures that would normally absorb or scatter light such as the bilipid layers that comprise cell membranes are rendered transparent.  Dyes can be integrated into living tissue and used as indicators for the presence of molecules of interest or to measure the observable state of cellular processes such as changes in membrane potential. Of course the addition of foreign molecules alters the optical properties of the tissue limiting penetration depth.  Resolution is limited by light scattering and the resulting loss in penetration depth this causes.  Advances in molecular functional magnetic resonance imaging (fMRI) may ultimately allow us to combine the specificity of electrophysiological recording techniques with the noninvasiveness and whole-brain coverage of current fMRI technology.
%
% Slide #1: Four example technologies, one in each quadrant of a single slide: (a) 3-D volume imaged using the CLARITY tissue preparation method from Karl Diesseroth's lab; (b) gene expression map generated using in situ hybridization from the Allen Institute Mouse Brain Atlas; (c) movie showing calcium imaging of live zebrafish from HHMI Janelelia Farm; (d) image of dendrite showing proteomic labeling produced using array tomography from Steven Smith's lab.
%
\item medium-term options [2--5 years] ---
  Biological organisms demonstrate a wide variety of molecules that carry out cellular computing, sensing and signalling.  These biomolecular devices are several orders of magnitude more efficient than CMOS devices.  Efforts so far to harness biological machines to perform logic and arithmetic are hampered by the fact that biological circuits coerced into implementing these traditional computing operations are orders of magnitude slower their CMOS counterparts.  However, for those computations required for survival, natural selection has stumbled on some incredibly efficient solutions.  Bioengineers are compiling libraries of biomolecules found in nature that perform such environmentally-specialized computations.  It is often said that, if you need a specific functional component for manipulating molecular information, you just have to find an organism that requires that function and re-purpose it for your application.  This synthetic-biology approach may be our most expedient option for getting to the next level in the next 2-5 years.  Examples currently under development include using retroviruses such as ``tame'' variants of the rabies and HIV virus to trace circuits, using DNA polymerase, the enzyme responsible for DNA replication, to record electrical potentials, using DNA sequences to barcode cells and cell products and modern, high-speed genome sequencing technology to create a map of all the synaptic connections in sample of tissue.  While it seems likely there will be proof-of-concept demonstrations of such technologies in the next few years, it will be some time before they develop to a point where they can be applied routinely in animal studies, and longer before they can safely be used in human studies. 

% Slide #2: Two technology-concept graphics shown side-by-side: (a) the ``molecular ticker-tape'' graphic from work at George Church's Harvard Medical School lab, and (b) the ``sequencing the connectome'' graphic from work at Tony Zador's Cold Spring Harbor lab.
%
\item longer-term options [5--10 years] ---
  Some neuroscientists believe that the ability to observe neural activity at substantially higher spatial and temporal resolution than currently possible will lead to new discoveries and new approaches to neuroscience.  One approach to achieving this level of detail is to enlist bioengineers and nanotechnologists to develop nanoscale recording and reporting devices that can be collocated with targets of interest in the neural tissue.  There are number of challenges to achieving such technology. Moore's law and related predictions of progress in miniaturizing integrated circuits suggest that in the next five years or so we will be able to manufacture chips of roughly the same size as a single cell --- less than 10 $\mu$m --- providing on the order of 10,000 transistors.  We would also have to reduce their power requirements to less than 10 nw to have some chance of powering the devices and dissipating their waste heat without causing cellular damage.  Most semiconductors used in chips are toxic to cells and so we would have to develop alternative  technologies like silicon bicarbide or find better ways of chemically isolating existing technologies such as cadmium.  Perhaps the biggest challenge involves solving the related reporting problem: getting the information out of the brain; obvious approaches to utilizing existing RF or optical communication technologies do not scale to billions of nanoscale reporters.  Safe, scalable solutions in this arena will likely require fundamental advances in science and engineering to achieve. 
%
% Slide #3: Four quadrants: (a) plot showing transistor count for a fixed 100$\mu$m2 area from 2000 to 2020 as projected by Moore's law; (b) plot showing illustrating modern processor technology efficiency progress in terms of computations per watt; (c) graphic showing off Hitachi's ``powder'' RFID chip technology; (d) carbon nanotube radio technology [Rutherglen and Burke].
%
\end{itemize}
Each of these three planning horizons, 1--2, 2--5, 5--10 years, offers opportunity for investment.  In 1--2 years, incremental improvements in functional magnetic resonance will continue accelerate the study of cognitive disorders of development and aging. Less-invasive ultrasound-based technologies for stimulation and surgical intervention are poised to deliver new treatments for neurodegenerative disease.  In the 2--5 year time frame, advances in synthetic-biology, DNA sequencing and multi-cellular recording promise insights into the function of neural circuits involving thousands of cells. Such insights are likely to yield new approaches to efficient computing, improved implants and prosthetic devices, and methods of harnessing natural computation. In 5-10 years, nanoscale sensors and robotics devices and the development of nanoscale communication networks will revolutionize health care.  New modes of human-computer interaction will provide the basis for seamless integration of computing and communication technologies with our natural cognitive capacities. 

We won't have to wait 2, 5, or 10 years for these technologies to provide immediate value.  We are already seeing early advances in personalized medicine in terms of better retinal and cochlear implants, metered drug delivery, precisely targeted cancer treatment, and deep-brain-stimulation implants to relieve chronic depression, essential tremor and Parkinson's disease.  More cumbersome instruments like MRI are being used to provide data on our emotional and cognitive states that can be used to train inexpensive, wearable devices that respond to our moods and preferences.  These personal assistants will transform the entertainment business and early devices already seeing adoption in the gaming industry.  New techniques in synthetic neurobiology show promise in the optogenetic control of the thalamus to interrupt seizures due to cortical injury, and every advance in the technology of efficient hybrid photovoltaics, genetically engineered biofuels and lab-on-a-chip microassays helps to move us closer to the goal of being able to monitor the brain at unprecedented scale and precision. Smart money will be watching for opportunties in all of these technology areas. 

%% file: introducing.tex
\section{Introduction}
\label{section:introducing}

% http://www.nytimes.com/2013/03/19/science/bringing-a-virtual-brain-to-life.html

Recent announcements of funding for two major initiatives~\cite{AlivisatosetalNEURON-12,MarkrametalHBP-12} in brain science have raised expectations for accelerated progress in the field of neuroscience and related areas of medicine.  
Both initiatives are depending on the development of new technologies and scientific instruments to realize their ambitious goals. 
Existing, well-established technologies will initially serve to propel the science forward, and these incumbent technologies will no doubt evolve to address new questions and offer new capabilities. 
However, we believe that current technologies will fall short in scaling to provide an appropriately wide range of co-registered assays across the whole brains of awake behaving animals and, in particular, human subjects\footnote{%
  The EU funded Human Brain Project~\cite{MarkrametalHBP-12} (HBP) and the US funded Brain Research through Advancing Innovative Neurotechnologies (BRAIN) Initiative both motivate their research programs in terms of the potential benefit to human health and welfare.  That said, there will continue to be a great deal of valuable research on alternative animal models, including primates, rodents and fish, as well as simpler organisms such as flies and other invertebrates.  These models offer a variety of experimental options depending on the organism, including cloning, genetic engineering, and the application of alternative recording technologies, e.g., embryonic zebrafish are essentially transparent allowing neural imaging opportunities impossible in other organisms~\cite{AkerboometalNEUROSCIENCE-12}.}.
We anticipate the need for new instruments that record diverse indicators of neural structure and activity\footnote{%
  The term ``activity'' is often used to describe the goals of the BRAIN (Brain Research through Advancing Innovative Neurotechnologies) Initiative. Indeed, the originally proposed name for the effort was the Brain Activity Map Project~\cite{AlivisatosetalNEURON-12}. Unfortunately, the term is ambiguous even among neuroscientists and can be used refer to very different sorts of brain-related activity, including action potentials, metabolic processes, e.g. mitocondrial efficiency, gene expression and diffuse neuromodulation, e.g. dopamine release in the {\it{substantia nigra}}, etc.} 
with substantially greater spatial and temporal resolution while providing rich contextual information of the sort required to relate these indicators to behavior and clinically relevant outcomes. 
In particular, this rich contextual information will be critical in analyzing recordings from multiple subjects or the same subject at different times and under varying conditions. 
There are precedents~\cite{SunkinetalNUCLEIC-13,HawrylyczetalPL0S-11} demonstrating how such scaling might be accomplished by taking advantage of the accelerating returns from computers, robotics and miniaturization that have made possible cell phones and the web. 
In addition, given their inherent potential for scaling, the nascent fields of synthetic biology and nanotechnology offer considerable promise but also pose formidable engineering challenges that will likely delay their availability as practical instruments for experimentalists. 

If successful these instruments will produce a veritable tsunami of data well beyond the capacity of any existing computational infrastructure to cope; we will be buried in riches with no way to realize their value; even current recording technologies seriously tax existing industrial-scale computing infrastructure.
The hope is that the accelerating returns from Moore's law and related trends in computing and storage technology will enable us to keep pace with the new technologies if we expend effort in developing and integrating appropriate data acquisition and analysis technologies throughout the scientific enterprise.
The key is to automate as much as possible. One lesson from web-scale computing is that, if you aren't willing to automate an information processing task, then don't bother trying to carry it out by ``hand'', because soon enough someone else will have figured out how to automate it and they will run rings around you. 
The application of high-throughput methods adapted from genomics research coupled with new animal models and transgenic strains for drug screening is a good example of acceleration that leverages the scaling opportunities inherent in robotics, information processing technology and molecular biology. 

There is no one technology that we can count on to make progress over the next decade.  Practically speaking, the scientific community will have to move forward on a portfolio of promising technologies to ensure steady progress on short- and medium-term objectives while laying the technological foundations for tackling more ambitious goals. 
It should be feasible to engage the capital markets to accelerate progress and underwrite some of the development costs as the the technologies that drive the science will have important practical applications in communications and computing, medicine, and entertainment to name just a few of the relevant areas.
The primary objective of this technical report is to populate the portfolio of promising technologies and provide scientists, entrepreneurs and investors with a basic understanding of the challenges and opportunities in the field as they are likely to evolve over the next decade.  

% http://www.stanford.edu/class/cs379c/class_messages_listing/index.html#recording_versus_reporting_readout

We begin by making a distinction between recording and reporting. Conceptually every technology we discuss will consist of a {\it{recorder}} and a {\it{reporter}} component: 
\begin{itemize}
\item recorder --- think of a recording device that when you hit the record button converts energy from the microphone (sensor) into a semi-permanent record (physical encoding) of the signal for subsequent reuse; the notion of recorder combines the functions of sensing the target signal and encoding it a form suitable for transfer.
\item reporter --- think of a news reporter who seldom observes the actual events but rather collects first-hand accounts, writes her interpretation of events and  posts them to her editor in a remote office; the notion of reporter combines the functions of transcoding, perhaps compressing and transmitting recorded information.
\end{itemize}
Recording technologies sample chemical, electrical, magnetic, mechanical and optical signals to record voltages, proteins, tissue densities, molecular concentrations, expression levels, etc.  

There are many existing recording technologies we discuss in this report, but we focus our attention primarily on reporting as we view this as one of the primary bottlenecks limiting our ability to accelerate our understanding of the brain. 

We divide recording technologies into two broad classes depending on whether the coding and signal-transmission components are located externally, outside of the target tissue, or internally, within the target tissue and typically co-located with the reporting components:
\begin{itemize}
\item external --- an external energy source radiating in the electromagnetic or acoustic spectrum is used to illuminate reporter targets within the tissue and the reflected energy is collected and analyzed to read out measurements.
\item internal --- reporters are coupled to local transducers that pre-process and package measurements in a format suitable for transmission over a communication channel employed to transmit the coded signals to an external receiver.
\end{itemize}
Most external technologies involve some form of imaging broadly construed.  In addition to those technologies explicitly described as ``imaging'', e.g., magnetic resonance imaging (MRI) and ultrasound imaging, we also include many of the technologies that incorporate the suffixes ``scopy'' or ``graphy'' in their name, e.g., scanning electron microscopy (SEM), near-infrared spectroscopy (NIRS), photoacoustic tomography and electroencephalography (EEG).  External technologies have the advantage that most of the energy required to make the measurements, process the information and extract the data from the tissue can be supplied from sources outside of the tissue.  The most common disadvantages of these technologies concern limitations in penetration depth due to scattering and absorption of energy within the tissue. 

In contrast, internal technologies require a source of energy to power the implanted devices and a method of safely dissipating of the resulting waste heat.  In addition, internal technologies generally require additional computational machinery to process, package and route raw measurements.  In some cases, there are biomolecular computing solutions that are energy efficient and compatible with surrounding tissue.  However, for many relatively simple computations we take for granted in conventional computing hardware, the biological options are too slow or error prone.  Nanotechnologies based on either semiconductor technology or silicon-and-synthetic-biology hybrids hold promise if we can continue to drive down size and power and overcome problems with toxicity and potential interference with normal cell function.

%% file: imaging.tex
\section{Evolving Imaging Technologies}
\label{section:imaging}

% %%%%%%%%%%%%%%%%%%%%%%%%%%%%%%%%%%%%%%%%%%%%%%%%%%%%%%%%%%%%%%%%%%%%%%%%%%%%%%%%%%%%%%%%%%%%%%%%%%%%%%%%%%%% %

Many of the most powerful recording and reporting technologies currently available can be characterized as imaging broadly construed in which some portion of the acoustic or electromagnetic spectrum is used to illuminate a target tissue and the returning signal is analyzed to extract useful information about structure, e.g., cytoarchitectural details, and function, e.g., membrane potentials and spike-timing data. These relatively mature technologies largely finesse the problems relating to powering reporting devices and carrying out the required computations involved in signal processing, compression and transmission by performing all these functions externally to the imaged tissue. 

These technologies are of particular interest since they are in daily use in hospitals throughout the world and are approved for human studies, unlike many of the other technologies we will discuss that are either not so approved or approved only in highly restrictive cases in which the patient has no other recourse for treatment or diagnosis. Both of the brain initiatives mentioned as depending on the accelerated development of the technologies surveyed in this report invoke understanding the human brain and relieving human suffering as motivation for their public funding.  The imaging technologies discussed in this section are among those most likely to produce such outcomes in the near term. 

With few exceptions, the technologies discussed in this section are relatively mature, were invented and refined largely independently of their applications in neuroscience, have promising directions for improved performance and extended capability, and are well funded by the private sector due to their use in the health care industry, industrial materials science and chip manufacturing.  The physics of nuclear magnetic resonance (NMR) has given rise to a family of technologies that are familiar in modern medical practice but that are also commonly employed in scientific instruments used in many other fields.  They include magnetic resonance imaging (MRI), functional MRI (fMRI), and diffusion tensor (functional) MRI (DTI). 

% %%%%%%%%%%%%%%%%%%%%%%%%%%%%%%%%%%%%%%%%%%%%%%%%%%%%%%%%%%%%%%%%%%%%%%%%%%%%%%%%%%%%%%%%%%%%%%%%%%%%%%%%%%%% %

% http://en.wikipedia.org/wiki/Nuclear_magnetic_resonance_spectroscopy

fMRI works by measuring local changes in hemodynamic response (blood flow) that are roughly correlated with neural activity.  The most commonly measured signal --- blood-oxygen-level-dependent (BOLD) contrast~\cite{OgawaetalPNAS-90} --- serves as a rough proxy for neural activity averaged over rather large local populations of neurons and thus offers only limited spatial and temporal resolution.  Studies involving fMRI offer some of the most powerful insights into high-level human cognition and its pathologies to date.  Using fMRI scientists are able to decode patterns of activity involving cognitive functions identified with anatomically distinct areas of the brain such as auditory and visual cortex, and even implement a rudimentary sort of mind reading~\cite{MitchelletalSCIENCE-08,KayetalNATURE-08}. State of the art fMRI resolution is on the order of 1 second temporal and 5 mm spatial. 

% http://en.wikipedia.org/wiki/Functional_magnetic_resonance_imaging

The path from blood oxygen level to neural activity is anything but direct. Glucose plus oxygen yields energy in the form of ATP and waste products in the form of carbon dioxide and water\footnote{%
  Oxygen is required in {\it{glycolis}} in which a glucose molecule is broken down into two three-carbon pyruvate molecules yielding two ATP molecules in the process.}. 
This reaction takes place in mitochondria many of which are located in the synaptic knobs at the end of axons where most of the brain's metabolic budget is spent. 
We pay a high metabolic cost for processing information; it requires on the order of $10^4$ ATP molecules to transmit a single bit at a chemical synapse~\cite{LaughlinNATURE-98}.
Specifically, most of the ATP consumed in the brain is spent pumping ions across cell membranes to maintain and restore resting potential following an action potential~\cite{HarrisetalNEURON-12}.
Despite its undisputed scientific and diagnostic merit, hemodynamic response and the BOLD contrast signal in particular are difficult to measure and complicated to interpret as an indicator for neural activity~\cite{LecoqetalNEUROSCIENE-09,BuxtonetalNEUROIMAGING-04,MagistrettiandPellerinPTRS-1991}. 
Better statistical tests to overcome measurement noise and alternative contrast agents are being sought after as alternative indicators for neural activity.  Calcium-sensitive agents make MRI more sensitive to calcium concentrations and calcium ions play an important role as messengers for cellular signalling pathways in active neurons.  Calcium indicators being developed for functional MRI may open up new opportunities for functional molecular imaging of biological signaling networks in awake, behaving organisms~\cite{AtanasijevicetalPNAS-06}.

% http://en.wikipedia.org/wiki/Diffusion_tensor_imaging

DTI estimates the rate of water diffusion at each location (voxel) in the imaged tissue.  This enables us to identify white matter which consists mostly of glial cells and myelinated axons and comprises the major information pathways connecting different parts of the brain.  DTI is particularly useful in a clinical setting for diagnosing stroke and white-matter pathologies. With a spatial resolution on the order of 100 $\mu$m it has provided some of the most stunning macroscale images to date for the Human Connectome Project (HCP)~\cite{SpornsetalPLoS-05,Seung2012}.  HCP is tasked with understanding the micro- and macro-scale connections in the human brain.  The general area of study is generally referred to as {\it{connectomics}}.  We'll return to discuss microscale connectivity --- connections between individual neurons --- in Section~\ref{section:automating}.

% %%%%%%%%%%%%%%%%%%%%%%%%%%%%%%%%%%%%%%%%%%%%%%%%%%%%%%%%%%%%%%%%%%%%%%%%%%%%%%%%%%%%%%%%%%%%%%%%%%%%%%%%%%%% %

% http://en.wikipedia.org/wiki/Super_resolution_microscopy

There are a number of other imaging technologies that share the property they operate outside of the target tissue relying entirely on external sources of power.  These include specialised conventional optical microscopes capable of $\mu$m resolution, and variants of confocal microscopy that achieve resolution below the diffraction limit to achieve resolution approaching 0.1 $\mu$m using laser light sources and super-resolution techniques~\cite{HuangetalARB-09,SchermellehetalJCB-10}.  We will pay particular attention in the remainder of this report to the use of devices capable of sub micron resolution that are typically used in conjunction with various dyes and contrast agents to resolve the details of neurons including their dendritic and axonal processes and to identify the presence of proteins and other molecules of interest. 

% http://en.wikipedia.org/wiki/Two-photon_excitation_microscopy

Scanning electron microscopes (SEM) can resolve details as small as a few nanometers, and, while not able to image individual atoms as are transmission electron microscopes, they are able to image relatively large areas of tissue which makes them the tool of choice for tracing neural circuits in chemically stabilized (fixed) tissue samples.  Unlike MRI and ultrasound these technologies are primarily of use with fixed tissue, {\it{in vivo}} experiments or animal studies requiring invasive procedures not likely to ever be approved for human subjects.  Two-photon excitation microscopy~\cite{HelmchenandDenkNATURE-05} is a fluorescence imaging technique that allows tissue penetration up to about one millimeter in depth and is particularly useful in circuit tracing and calcium imaging~\cite{GreweetalNATURE-10} in living tissue and genomic, transcriptomic, and proteomic maps of fixed neural tissue.

% %%%%%%%%%%%%%%%%%%%%%%%%%%%%%%%%%%%%%%%%%%%%%%%%%%%%%%%%%%%%%%%%%%%%%%%%%%%%%%%%%%%%%%%%%%%%%%%%%%%%%%%%%%%% %

% http://en.wikipedia.org/wiki/Positron_emission_tomography

There are other imaging technologies relying on light or more exotic physics that deserve brief mention.  Near-infrared spectroscopy (NIRS) is of interest as a non-invasive technology that can be used to detect a signal similar to BOLD but without an expensive and cumbersome MRI magnet.  NIRS provides limited spatial resolution and depth of field but is already proving useful as an inexpensive sensor in building brain-computer interfaces (BCI) for gaming headsets.  Positron emission tomography (PET) provides some of the same capabilities as MRI but has received less attention in part due to its associated radiation risk and competing technologies maturing to subsume its most clinically relevant capabilities. 

% http://en.wikipedia.org/wiki/Electroencephalography

There is also a set of technologies that attempt to directly sense electrical signals resulting from neural activity without penetrating brain tissue and causing cellular damage. Electroencephalography (EEG) is perhaps best known of these technologies. EEG measurements can be recorded from awake behaving humans with no surgical intervention, but because the electrodes are placed on the skin covering the skull it offers a spatial resolution on the order of 5cm, and, while the temporal resolution can be as high 1KHz, measurements average over large populations of neurons.  Electrocorticography (ECoG) allow us to measure local field potentials (LFPs) with temporal resolution of approximately 5 ms and a spatial resolution of 1 cm but involves a craniotomy as it requires a grid of electrodes placed directly on the exposed surface of the brain.  Advanced ECoG methods have successfully decoded speech from patterns of activity recorder over auditory cortex~\cite{PasleyetalPLoS-12}.

% http://en.wikipedia.org/wiki/Magnetoencephalography

Magnetoencephalography (MEG) --- like EEG and ECoG --- measures the net effect of currents flowing through the dendrites of neurons during synaptic transmission --- not, it is worth noting, action potentials.  The synchronized currents of neurons generate weak magnetic fields just barely detectable above the background levels even with extensive shielding and superconducting magnetometers.  On the order of 50,000 active neurons are required to generate a detectable signal.  MEG has the temporal resolution of EEG and the spatial resolution of ECoG, however unlike ECoG MEG does not require a craniotomy. 

% %%%%%%%%%%%%%%%%%%%%%%%%%%%%%%%%%%%%%%%%%%%%%%%%%%%%%%%%%%%%%%%%%%%%%%%%%%%%%%%%%%%%%%%%%%%%%%%%%%%%%%%%%%%% %

Light in the visible range of the electromagnetic spectrum cannot penetrate deeply into tissue.  Clearing reagents like CLARITY~\cite{ChungetalNATURE-13} and Clear$^{\mbox{\rm{T}}}$~\cite{KuwajimaetalDEVELOPMENT-13} can be used to prepare tissue in such a way that structures that would normally absorb or scatter light such as the bilipid layers that comprise cell membranes are rendered transparent, but this approach doesn't apply to living tissue.  At radio frequencies below 4 MHz the body is essentially transparent to the energy which makes it a candidate for transmitting data but not imaging.  Light in the near-infrared range with a wavelength of about 800 nm to 2500 nm can penetrate tissue to several centimeters and is used in both imaging and stimulating neural tissue.  However, the electromagnetic spectrum is not the only alternative we have for non-invasively probing the brain.

% http://en.wikipedia.org/wiki/Acoustic_microscopy

Ultrasound pressure waves are routinely used in diagnostic medical imaging and easily penetrate tissue to provide real-time images of the cardiovascular system.  The ultrasonic frequencies used for diagnostic imaging range typically between 2 MHz and 20 MHz.  Spatial resolution on the order of 1~$\mu$m is possible with frequencies in the 1-2 GHz range, but attenuation in soft tissue increases as a function of the frequency thereby reducing penetration depth\footnote{%
  Theoretically, attenuation increases with the square of the frequency, but linear increases have been reported for several biological tissues~\cite{Hoskinsetal2010}. Unfortunately, the empirically-derived attenuation coefficient for neural tissue is 200 times that of water --- 0.435 versus 0.002 (db cm$^{-1}$ MHz$^{-1}$)~\cite{Hoskinsetal2010} --- despite the two materials having similar density, 1.0 versus 1.02 (g/cm$^3$), speed of sound, 1480 versus 1550 (m/sec) and acoustic impedance, 1.48 versus 1.60 ([kg/(sec m$^2$)] $\times$ 10$^6$).}.
This loss of penetration depth can be compensated somewhat by increasing the signal intensity while limiting exposure.  For exposures longer than ten seconds, intensity levels less than 0.1 (W/cm$^2$) are generally deemed safe for diagnostic imaging.  For a 70 MHz signal traveling through water, the attenuation coefficient is 10 (dB/cm) and, given an intensity of 0.1 (W/cm$^2$), the maximum effective depth would be in the range 1.5-2 cm.\footnote{%
  The formula for the attenuation coefficient in water as a function of frequency is $\alpha = 2.17 \times 10^{-15} \times$ {\it{f}}$^{\mbox{~2}}$ (dB/cm) where $f$ is the frequency~\cite{MordfinASTM-02}.  In order to achieve a spatial resolution of 15 $\mu$m, we would need a frequency of 100 MHz and given that the corresponding $\alpha$ is approximately 20 (dB/cm) it would be difficult if not impossible to safely penetrate to the maximum depth required to image an entire human brain.}.

Ignoring temperature and barometric pressure, ultrasound travels about five times faster in water (about 1500 m/s) than it does in air (about 300 m/s). Most ultrasound technologies exploit the piezoelectric effect for acoustic signal transmission, reception or both.  When materials like ceramics, bone and DNA are mechanically stressed, they accumulate an electrical charge. Conversely, by applying an external electric field to such materials, we can induce a change in their static dimensions.  This inverse piezoelectric effect is used in the production of ultrasonic sound waves.

Focused ultrasound (FUS) technologies developed for medical applications employ a phased-array of piezoelectric transducers to produce multiple pressure waves whose phase is adjusted by introducing delays in the electrical pulses that generate the pressure waves.  By coordinating these delays, the focal point --- point of highest pressure and thus highest temperature in the tissue --- can be precisely controlled to avoid cell damage.  FUS can be used for deep brain stimulation and has demonstrated promise in clinical trials on patients suffering from essential tremor.  FUS has also been used to alter the permeability of the blood-brain barrier to allow the controlled diffusion of drugs or other nanoparticles across the blood-brain barrier~\cite{WangetalJBO-12}.  A hybrid near-infrared-plus-ultrasound-imaging technology called {\it{photoacoustic spectroscopy}} has been shown effective in monitoring focused-ultrasound-induced blood-brain barrier opening in a rat model {\it{in vivo}}.  

High-intensity focused ultrasound (HIFU) can be used to destroy a tumor in the breast, brain or other tissue without cutting into the surrounding tissue.  In the case of the brain, the cranium poses a challenge due to its variable thickness, but this can be overcome either by performing a craniotomy or by using a CT scan to construct a 3-D model of the skull and then generating a protocol based on this model that adjusts the delays to correct for aberrations in signal propagation due to the changes in thickness of this particular skull~\cite{VlijmenetalCOI-11}.  HIFU offers an alternative to Gamma-knife surgery without the attendant radiation risk though there are some drawbacks due to the fact that ultrasound waves, unlike ionizing radiation, can be deflected and scattered.

% http://en.wikipedia.org/wiki/Acoustic_microscopy 

Ultrasound technologies like transcranial doppler (TCD) imaging have been used to measure the velocity of blood flow through the brain's blood vessels, and provide fast, inexpensive diagnostics for brain injuries~\cite{BeckerandBergMD-01}. TCD offers temporal resolution comparable to other neuroimaging techniques, but spatial resolution is relatively poor at the frequencies typically used in diagnostic imaging.  High-frequency acoustic microscopes do exist, however, and are routinely employed in clinical settings, most notably in producing {\it{sonograms}} used in ophthalmology for treating glaucoma~\cite{SilvermanCEO-09}.  For the time being, it seems the most likely applications of ultrasound in experimental neuroscience will involve diffuse stimulation using nanobubble contrast agents~\cite{DijkmansetalEJE-04} and manipulation of the blood-brain-barrier for introducing engineered biomolecules and nanoparticles into the brain --- see Appendix~\ref{appendix:anjali}. 

% %%%%%%%%%%%%%%%%%%%%%%%%%%%%%%%%%%%%%%%%%%%%%%%%%%%%%%%%%%%%%%%%%%%%%%%%%%%%%%%%%%%%%%%%%%%%%%%%%%%%%%%%%%%% %

The technologies discussed in this section are relatively mature.  In some cases, they have run up against fundamental physical limitations, and improvements in resolution and accuracy are likely to come from incremental engineering.  That said they are also the incumbents in the extant technology race; their technology has set a high bar, and even incremental improvements will likely prove sufficient to maintain their dominance and market share.  In many cases, they also have the advantage that the medical profession is conservative and reluctant to abandon technology in which they have invested time learning how to master.  To break into these established markets, new companies will have to demonstrate substantially new capabilities to attract funding and successfully launch products.  

Attractive capabilities that might serve as game changers include technologies that combine recording and stimulating, e.g., thus enabling the use of feedback for direct control of treatment, multi-function technologies that monitor several indicators at once, e.g., recording co-located electrical and proteomic activity, lighter, smaller, less intrusive technologies, e.g., wearable or implantable devices to support remote monitoring and improve patient compliance, and technologies that significantly expedite existing protocols or eliminate them entirely, e.g., portable, affordable alternatives to diagnostic MEG, MRI and PET for small family medical practices.

Speed definitely matters in both clinical and scientific studies.  Multi-plane, parallel-imaging NMR technologies\footnote{%
  Parallel acquisition techniques combine the signals of several coil elements in a phased array to reconstruct the image, the chief objective being either to improve the signal-to-noise ratio or to accelerate acquisition and reduce scan time.}
have accelerated scanning four-fold, and there are likely greater gains to be had as other parts of the processing pipeline catch up.  With the advent of portable devices for bedside ultrasound scanners and a move by manufacturers to support beam-forming in software researchers have been able to improve throughput thirty-fold with no reduction in resolution~\cite{MadoreetalULTRASONICS-09}.  The latest SEM technology promises high-throughput, large-area imaging using a multi-beam scanning.  The electron source is split into multiple beams --- as many as sixty --- and all of the beams are scanned and recorded from in parallel resulting in a sixty-fold speedup in acquisition.

While some improvements among the incumbent technologies will require fundamental advances in science, improvements requiring computation can immediately take advantage of the accelerating returns from advances in computer design and chip fabrication.  For example, new signal-processing algorithms running on faster hardware can increase accuracy with no loss in throughput by sampling more frequently, sampling more intelligently or using existing sampling strategies but spending more time analyzing the samples.  MRI, ultrasound imaging and scanning electron microscopy are all poised to make substantial improvements in throughput by exploiting computation to accelerate the acquisition and analysis of data.

% %%%%%%%%%%%%%%%%%%%%%%%%%%%%%%%%%%%%%%%%%%%%%%%%%%%%%%%%%%%%%%%%%%%%%%%%%%%%%%%%%%%%%%%%%%%%%%%%%%%%%%%%%%%% %

%% file: probing.tex
\section{Macroscale Reporting Devices}
\label{section:probing}

% %%%%%%%%%%%%%%%%%%%%%%%%%%%%%%%%%%%%%%%%%%%%%%%%%%%%%%%%%%%%%%%%%%%%%%%%%%%%%%%%%%%%%%%%%%%%%%%%%%%%%%%%%%%% %

In this section we continue our discussion of relatively mature technologies by examining tools for reporting on the microscale properties of individual cells using technologies whose components are implemented at the macroscale.  In Sections~\ref{section:sequencing} and~\ref{section:shrinking} we will return to the problem of resolving details at the microscale but this time employing microscale components based on, respectively, re-purposed biological components and hybrid technologies that combine biological parts with inorganic devices including novel applications of semiconductor and integrated-circuit technology. 

% %%%%%%%%%%%%%%%%%%%%%%%%%%%%%%%%%%%%%%%%%%%%%%%%%%%%%%%%%%%%%%%%%%%%%%%%%%%%%%%%%%%%%%%%%%%%%%%%%%%%%%%%%%%% %

% http://en.wikipedia.org/wiki/Electrophysiology

The classic single-probe electrode used in electrophysiology consists of a solid conductor in the form of a needle that can be inserted into the tissue to record the local field potentials resulting from the activation of neurons in close proximity to the uninsulated tip of the electrode.  In Section~\ref{section:automating} we review current methods of automating the insertion and manipulation of such devices thereby eliminating the most time- and labor-intensive part of their application in the lab.

% http://en.wikipedia.org/wiki/Multielectrode_array

While still used in practice many scientists now employ multi-electrode arrays consisting of hundreds of electrodes arranged in a grid in order to better resolve the activity of individual neurons.  These arrays can be implanted in living tissue for {\it{in vivo}} experiments or cells can be cultured on arrays for {\it{in vitro}} studies. Arrays constructed from stainless steel or tungsten are now being replaced by silicon-based devices that can benefit from the same technologies used in chip manufacture, and technologies originally intended for the lab are being refined for use as permanent implants in humans spurring innovation in the design of biocompatible devices. 

% http://en.wikipedia.org/wiki/Tetrode_(biology)

Tightly grouped bundles of very small electrodes are often used to enable more accurate local readings.  Four such electrodes in an arrangement called a {\it{tetrode}} provide four spatially-distributed channels that can be used to better separate the signals originating from different neurons.  The basic idea can extended and {\it{polytrodes}} consisting of 54-channel high-density silicon electrode arrays have been used to make simultaneous recordings from more than 100 well-isolated neurons~\cite{BlanhceetalJoN-05}.  Flat arrays called {\it{microstrips}} with as many as 512 electrodes and 5 $\mu$m spacing have been used to record from and stimulate cells in the retina and visual cortex~\cite{LitkeNIMP-98}.  

% %%%%%%%%%%%%%%%%%%%%%%%%%%%%%%%%%%%%%%%%%%%%%%%%%%%%%%%%%%%%%%%%%%%%%%%%%%%%%%%%%%%%%%%%%%%%%%%%%%%%%%%%%%%% %

The recording tips of the electrodes in the multi-electrode arrays mentioned above are typically arranged to lie on the same plane and so these arrays are essentially 2-D probes.  To enable an additional degree of freedom in recording, these probes can be advanced or retracted in small steps to sample from a 3-D volume, but this implies that at every point in time the samples are all drawn from a planar region. It is worth noting that while the tools of electrophysiology are used primarily for sampling extracellular voltages, electrophysiology can also measure internal voltages of a single neurons. In studying hippocampal place cell activity, sub-threshold dynamics seem to play an important role and there are currently limited ways to get a similar signal with the other technologies~\cite{HarveyetalNATURE-09}.

Single probes with multiple recording sites along their length have been developed with as many as 64 channels~\cite{DueetalPLoS-11}.  These probes are particularly revealing when inserted into cortical tissue either vertically or horizontally relative to the cortical surface to record from different types of neurons located in multiple layers or from multiple neurons of the same type of neuron located within the same layer.  If you ask an experimental neuroscientist interested in early vision if it would help to record more densely from visual cortex, the answer would likely be yes since it is a common and puzzling experience having recording from a neuron that appears to strongly respond to one stimuli, only to find when you move your probe scant microns you encounter a neuron that responds strongly to a completely different stimuli. 

There is also work developing true 3-D arrays using flexible materials for use in chronic implants~\cite{ChenetalMEMS-10}.  Optogenetics has opened the door to implants that can exert exquisite control over individual neurons with optical-fibers being the method of choice for delivering light to exert such control.  In principle the same wave guides used to deliver light to activate neurons can be multiplexed to receive light from biomolecules designed to record from neurons. Optical wave guides that can be used for delivering light to excite or inhibit individual neurons in a 3-D volume have been demonstrated~\cite{ZorzosetalOPTICS-12} and since these probes were fabricated using CMOS technology the expectation is that these technology will scale to thousands of targets with the 3-D volume.

% %%%%%%%%%%%%%%%%%%%%%%%%%%%%%%%%%%%%%%%%%%%%%%%%%%%%%%%%%%%%%%%%%%%%%%%%%%%%%%%%%%%%%%%%%%%%%%%%%%%%%%%%%%%% %

% http://en.wikipedia.org/wiki/Calcium_imaging

A {\it{fluorophore}} is a fluorescent compound that re-emits photons upon excitation and can be covalently bonded to a macromolecule to serve as a dye, tag or marker.  Fluorophores can be employed as microscale recorders capable of resolving nanoscale details. The method of {\it{calcium imaging}} is perhaps the most successful such application of this idea to recording from many neurons simultaneously~\cite{GrienbergerandKonnerthNEURON-12}.  The synaptic transmission of action potentials is controlled in part by an influx of calcium into the synaptic terminal of transmitting neuron's axon\footnote{%
  There is also post-synaptic (somal) calcium influx during depolarization, and for this reason the method of calcium imaging can be used to analyze the antecedents to action potentials in firing neurons not just the consequences manifest in their axonal processes.}.
The distribution of calcium in synapses can be used as a proxy for neural activity.  Detection is accomplished using genetically encoded calcium indicators (GECI) that respond to the binding of Ca$2^{+}$ ions by changing their fluorescence properties~\cite{AkerboometalNEUROSCIENCE-12}. 

Imaging is typically accomplished using one- or two-photon microscopy but it was recently shown to be possible to image the entire brain of a larval zebrafish at 0.8 Hz and single-cell resolution using laser-scanning light-sheet microscopy~\cite{AhrensetalNATURE-13}, and, despite calcium being a lagging indicator, it is possible to employ calcium imaging data to reconstruct spike trains using Monte Carlo sampling and super-resolution techniques~\cite{VogelsteinetalBJ-09}.  Miniaturized fluorescence microscopes offering~\textasciitilde{}0.5 mm$^2$ field of view and~\textasciitilde{}2.5 micron lateral resolution enable researchers to record from awake, behaving mice~\cite{GhoshetalNATURE-11}.  

% http://en.wikipedia.org/wiki/Patch_clamp

Patch clamping is a bench technique for recording from single or multiple ion channels in individual neurons. It is the most precise method available to the neuroscientist for recording electrical activity in neurons.  Until recently it required a highly skilled technician to carry out the procedure.  The method has now been automated~\cite{KodandaramaiahetalNATURE-12} opening up the possibility of highly-parallel, robotically-controlled experiments.  It is possible to build nanoscale recorders that indicate whether a specific ion channel is open or closed, and so it is natural to ask if one could not simply record from {\it{all}} the ion channels on an axon and use this information to reconstruct the propagation of action potentials.  While conceivable in principle, it is not likely to prove practical any time soon due to the simple fact that there on the order of 1,000 sodium pumps (voltage-gated ion channels) per $\mu{m}^2$ of axonal membrane surface or about a million sodium pumps for a small neuron.

% %%%%%%%%%%%%%%%%%%%%%%%%%%%%%%%%%%%%%%%%%%%%%%%%%%%%%%%%%%%%%%%%%%%%%%%%%%%%%%%%%%%%%%%%%%%%%%%%%%%%%%%%%%%% %

There are several relatively static pieces of information that neuroscientists would like to have for their experimental subjects: the set of all RNA molecules transcribed in each cell type --- referred to as the {\it{transcriptome}}, the set of all proteins expressed by each cell type --- called the {\it{proteome}}, and a complete map of the connections between cells along with some measure of the strength of those connections referred to as the {\it{connectome}}\footnote{%
  If exhaustive, the connectome would have to include axodendritic, axosomatic and dendrodentritic connections along with a classification of the pre- and post-synaptic cell types, relevant neurotransmitters, and information on whether the connections are excitatory or inhibitory.}.
In terms of basic genomic information, we might sequence an instance of each cell type or multiple instances of the same cell type drawn from different locations in order to search for epigenetic differences.  However, full sequencing is likely unnecessary given that we know what to look for and so, for practical purposes, it should suffice to apply a method such as {\it{in situ}} hybridization or immunoflorescence to obtain genetic signatures for cells or create a map relating genetic differences to locations corresponding to the coordinates of small volumes of tissue.

One point to emphasize is the importance of maps and why they are so useful to scientists. In particular, we envision multiple maps: genomic, connectomic, transcriptomic, and proteomic maps along with sundry other maps registering activity and metabolic markers. 
These maps will be significantly more useful if different maps from the same subject can be registered with one another so we can relate their information spatially --- and temporally in the case of maps with a temporal dimension. 
Equally important is the ability to relate maps from different subjects of the same species, as in the case of comparing healthy and diseased brains.  As simple as this sounds, it is an enormously complex problem.  That the Allen Institute was able co-register the maps of almost identical cloned mice in building mouse-brain atlas~\cite{SunkinetalNUCLEIC-13} is a significant accomplishment and preview of some of the difficulties likely to be encountered in building a human-brain atlas. 
Ideally we would like a standardized method for moving between organisms; such methods exist but terminology differences between humans and model organisms complicate the mapping.  A controlled ontology analogous to the NIH-maintained Medical Subject Headings (MeSH) for neuron cell types, neuroanatomical landmarks, etc.\ would enable more fluid transitions between different experimental results.

% %%%%%%%%%%%%%%%%%%%%%%%%%%%%%%%%%%%%%%%%%%%%%%%%%%%%%%%%%%%%%%%%%%%%%%%%%%%%%%%%%%%%%%%%%%%%%%%%%%%%%%%%%%%% %

% http://en.wikipedia.org/wiki/Connectomics

Though there are many variations the most common approach to generating a connectome involves stabilizing and staining the tissue sample, slicing it into thin sections, scanning each slice with an electron microscope, and then applying some method to segment the cell bodies and identify connections between cells~\cite{BriggmanetalNATURE-11,MikulaetalNATURE-12}.  This last stage is the most time consuming by several orders of magnitude as it typically involves the intervention of a human expert.  While automating cell-body segmentation is still an open problem, most computer vision experts believe that it will be solved by some combination of better algorithms, better stains and better imaging hardware~\cite{JainetalCON-10}. Refinements of tissue-preparation technologies such as CLARITY~\cite{ChungetalNATURE-13} and Clear$^{\mbox{\rm{T}}}$~\cite{KuwajimaetalDEVELOPMENT-13} --- see Section~\ref{section:imaging} --- may eventually prove useful in segmenting cell bodies.  We will return briefly to the problem automating the production of connectomes in Section~\ref{section:automating}.  

% http://en.wikipedia.org/wiki/Proteomics

Array tomography is a method for extracting a proteomic map which is superficially similar to the method described above for producing connectomes in all but the last step.  Instead of applying a computer vision algorithm to segment each SEM image, you apply the method of immunofluorescence to tag proteins with fluorescent dyes so they can be identified with a scanning electron microscope~\cite{MichevaandSmithNEURON-07}.  Alternative methods of tagging such as fluorescent {\it{in situ}} sequencing (FISSEQ) on polymerase colonies~\cite{MitraAB-03} may offer a way around current limitations in the number of tags --- and hence proteins --- you can distinguish between in a single tagging-and-imaging pass over a given volume. 

% http://en.wikipedia.org/wiki/Immunofluorescence

A volumetric transcriptomic map identifies for each 3-D volume in a tissue sample the set of all RNA molecules produced in the population of cells located within the volume.  Ideally such a map would have a temporal extent given that the type and number of transcribed genes will vary over time, at the very least over the development of the organism~\cite{KangetalNATURE-11}.  The simplest approach to constructing a static volumetric transcriptomic map is to slice the sample into small volumes, separate out the RNA and sequence using RNA microarrays, FISSEQ or some other high-throughput sequencing method. 

% %%%%%%%%%%%%%%%%%%%%%%%%%%%%%%%%%%%%%%%%%%%%%%%%%%%%%%%%%%%%%%%%%%%%%%%%%%%%%%%%%%%%%%%%%%%%%%%%%%%%%%%%%%%% %

In this section, we looked at reporting technologies consisting of macroscale components which, while capable of reporting out the information provided by microscale recorders, are volumetrically limited by absorption and scattering.  The exception being static maps of tissue samples that are dissembled for analysis assuming that the organism can be sacrificed.  By employing a fiber-optically coupled microendoscope~\cite{BarrettoandSchnitzerCSHP-12} it is possible to image cells deep within the brains of live animals over extended periods, but with reduced field of view and the invasive introduction of the endoscopic device.  Dynamic maps that record neural activity in awake behaving humans are limited in scale by existing reporter technology.  In Sections~\ref{section:sequencing} and~\ref{section:shrinking}, we will explore technologies for implanting microscale reporter co-located with their microscale recorder counterparts to enable scaling activity maps to even larger spatially distributed ensembles of neurons. 

% %%%%%%%%%%%%%%%%%%%%%%%%%%%%%%%%%%%%%%%%%%%%%%%%%%%%%%%%%%%%%%%%%%%%%%%%%%%%%%%%%%%%%%%%%%%%%%%%%%%%%%%%%%%% %

%% file: automating.tex
\section{Automating Systems Neuroscience}
\label{section:automating}

% %%%%%%%%%%%%%%%%%%%%%%%%%%%%%%%%%%%%%%%%%%%%%%%%%%%%%%%%%%%%%%%%%%%%%%%%%%%%%%%%%%%%%%%%%%%%%%%%%%%%%%%%%%%% %

The notion of testable hypothesis and the quest for simple, elegant explanatory theories is at the very foundation of science.  The idea of automating the process of hypothesis generation, experimental design and data analysis, while once considered heretical, is now gaining favor in many disciplines, particularly those faced with explaining complex phenomena.  The search for general principles --- the holy grail of modern science --- sounds so reasonable until you ask scientists what would qualify for such a principle, and here you are likely to get very different answers depending on whom you ask.  We would like the world to be simple enough accommodate such theories, but there is no reason to expect nature will cooperate.  Perhaps in studying complex systems like the brain, we'll have to settle for a different sort of comprehension that speaks to the emergent properties and probabilistic interactions among simpler components and mathematical characterizations of their equilibrium states~\cite{AlivisatosetalNEURON-12,Raoetal2002,HopfieldPNAS-82}.

The term ``emergent'' is often used derisively to imply inexplicable, opaque or even mystical.  However, there is reason to believe at least some of the processes governing the behavior of neural ensembles are best understood in such terms~\cite{Strogatz2002,Buzsaki2006}.  This is not a call for abandoning existing theories or the search for general principles, but rather a suggestion that we consider new criteria for what constitutes an adequate explanation, entertain new partnerships with computational scientists and where possible automate aspects of the search for knowledge, and explore new classes of theory that demand data and computation to discover and evaluate.  Computation- and data-driven approaches don't necessarily imply recording from millions of neurons, though techniques from data mining and machine learning may be the best tools for making sense of such data.  Indeed, there already exist powerful tools in the current repertoire --- see Sections~\ref{section:imaging} and~\ref{section:probing} --- that can be applied to produce data of sufficient quantity and quality to explore interesting such phenomena~\cite{BorgersetalPLOS-12,CanoltyetalPNAS-10,SpornsetalPLoS-05}.  Distributed and sparse coding models of visual memory were initially motivated by computational and statistical arguments~\cite{Barlow61,OlshausenandFieldNC-96} and subsequent work developing probabilistic models crucially depend on analyses that would not have been possible without modern computational tools and resources~\cite{SerenoandLehkyFCR-10,CanoltyetalPNAS-10,RozelletalNC-08}.

% %%%%%%%%%%%%%%%%%%%%%%%%%%%%%%%%%%%%%%%%%%%%%%%%%%%%%%%%%%%%%%%%%%%%%%%%%%%%%%%%%%%%%%%%%%%%%%%%%%%%%%%%%%%% %

% http://neuron.duke.edu/userman/2/pioneer.html, http://en.wikipedia.org/wiki/Hodgkin-Huxley_model

As our understanding of smaller circuits and diverse molecular, electrical and genetic pathways improves we can expect to see increased reliance on high-fidelity simulations and data-driven model selection\footnote{In statistics and machine learning, {\it{model selection}} is the problem of picking from among a set of mathematical models all of which purport to describe the same data set.  The task can also involve the design of experiments to ensure that the data collected is well-suited to the problem of model selection.}~\cite{StougiannisetalSIGMOD-13,TauheedetalICDE-12}.  These approaches will enjoy accelerated returns from advances in computer technology allowing scientists to explore and model larger ensembles. In the past, mathematicians, statisticians and computational scientists interested in analyzing neural data often found themselves isolated at the far end of a pipeline, shut out of the preliminary discussions involved in designing experiments and vetting theories, and privy to such earlier decisions only through the spare, stylized format of academic publication.  This sort of specialization made it difficult to approach the problem of model selection in a systematic end-to-end fashion with opportunities to adjust each step of the process from stipulating which measurements are taken and how the data is annotated, curated and stored electronically to defining which hypotheses and classes of models are appropriate in explaining the data. Today computational scientists are integral members of multidisciplinary teams. 

The idea of including computational scientists early in designing experiments, vetting theories and creating models and frameworks for understanding is not new.  Neural simulators have been around as long as computers\footnote{In developing the Hodgkin-Huxley model of action potentials~\cite{HodgkinandHuxleyJoP-52}, Huxley carried out extensive calculations on a manually-cranked calculator to make predictions about how action potentials would change as a function of the concentration of extracellular sodium.  His predictions were later confirmed by Hodgkin's experiments with giant squid axons~\cite{HodgkinJoP-76}. Around the same time, a model of synaptic plasticity developed by Donald Hebb~\cite{Hebb49} was simulated on an early digital computer at MIT by Farley and Clark in 1954~\cite{FarleyandClarkTIC-54}.}, and modern high-fidelity, Monte Carlo simulations~\cite{KerretalSCIAM-08} are capable of modeling the diffusion and chemical reactions of molecules in 3-D reconstructions of neural tissue and are used for a wide range of {\it{in silico}} experiments, e.g., providing evidence for ectopic neurotransmission in synapses~\cite{CogganetalSCIENCE-05} and accurate simulations of 3-D reconstructions of synapses yielding new insights into synaptic variability~\cite{StilesetalMCELL-01}.

While several large-scale simulations~\cite{AnanthanarayananetalSC-09,IzhikevichandEdelmanPNAS-08,MarkramNATURE-06} have received attention in recent years, skeptics believe that such efforts are premature given the present state of our knowledge~\cite{WaldropNATURE-12}.  Whether true or not, many researchers characterise their work as trying to understand how neural circuits give rise to behavior.  But the gap between circuits and behavior is wide, and an intermediate level of understanding might be achieved by characterizing the neural computations which occur in populations of neurons~\cite{CarandiniNATURE-12}.  In the same way that knowing the primitive operators in a programming language is essential to understanding a program written in that language, so too the computations we use to characterize the function of smaller circuits might eventually provide us with a language in which to describe the behaviors supported by circuits comprised of larger ensembles of neurons.  

In some cases, it is instructive to discover that one sort of information, say spiking behavior, can be recovered from another sort of information, say calcium influx, using an appropriate machine learning algorithm. Computational neuroscientists demonstrated that accurate spike-timing data could be recovered from the calcium imaging of neural populations opening the possibility of recording from neural circuits and inferring spikes --- the gold standard for summarizing the electrical behavior of individual neurons --- without patch clamping or inserting electrodes~\cite{VogelsteinetalBJ-09}.  In developing the Allen Mouse Brain Atlas~\cite{SunkinetalNUCLEIC-13}, the Allen team had to invent techniques for registering neural tissue samples from multiple cloned mice against standardized anatomical reference maps and, in the process, they developed powerful new tools for visualizing 3-D connectivity maps with detailed gene-expression overlays.  Patterns apparent from gene expression maps are often as useful if not more so than canonical reference maps for identifying functional areas, and this observation may be of crucial importance when we attempt to build a human brain atlas. 

It is worth mentioning that the Allen Mouse Brain Atlas required the collaboration of scientists from diverse fields including neuroanatomy, genomics and electron microscopy and would not have been possible without the involvement of mathematicians and computer scientists from the very start of the project.  Given the trend, we expect more successful collaborations involving neuroscientists and biophysicists with a deep understanding and broad interpretation of computation as manifest in biological systems and computer scientists and electrical engineers with extensive knowledge and appreciation of biological organisms and the many chemical, electrical, mechanical and genetic pathways that link those organisms to the environments in which they evolved.  And in, terms of leveraging tools, techniques and technologies that benefit from the accelerating returns of computation, we can also expect help from another unexpected quarter: industrial and consumer robotics.

% %%%%%%%%%%%%%%%%%%%%%%%%%%%%%%%%%%%%%%%%%%%%%%%%%%%%%%%%%%%%%%%%%%%%%%%%%%%%%%%%%%%%%%%%%%%%%%%%%%%%%%%%%%%% %

% http://www.wormatlas.org/EMmethods/ATUM.htm, http://en.wikipedia.org/wiki/Serial_block-face_scanning_electron_microscopy

The automation of serial sectioning, section handling and SEM imaging\footnote{As discussed in Section~\ref{section:probing}, serial-section block-face scanning electron microscopy is one of the primary methods used in connectomics and proteomics for creating 3-D maps of stabilized, stained neural tissue~\cite{BriggmanandBockCON-12}.} for large-volume 3-D reconstruction and array tomography has enormously sped up the collection of data required for applications in connectomics and proteomics.  Developments like the automatic tape-collecting ultra-microtome (ATUM) remove the requirement for a skilled human in the loop and dramatically reduce sorting errors and tissue-handling damage~\cite{SchaleketalMM-11}.  The automated 3-D histological analysis of neural tissue is still in the early stages of development but a great deal of progress has been made~\cite{MishchenkoJNM-09} and we can expect rapid improvement in the next few years fueled by advances in machine learning~\cite{JainetalCON-10} and high-performance computing.

% %%%%%%%%%%%%%%%%%%%%%%%%%%%%%%%%%%%%%%%%%%%%%%%%%%%%%%%%%%%%%%%%%%%%%%%%%%%%%%%%%%%%%%%%%%%%%%%%%%%%%%%%%%%% %

Indeed, there are many repetitive tasks routines carried out in the lab that can be automated by machine learning and computer vision or accelerated by robotically controlled instruments.  Even such delicate tasks as patch-clamp electrophysiology~\cite{KodandaramaiahetalNATURE-12} and multi-electrode array insertion~\cite{ZoccolanetalPNAS-09} can be performed by programmable robots.  Increasingly the manufacturers of scientific instruments are replacing hardware components by software components so they can offer performance enhancements by simply upgrading the software or swapping out a circuit board and replacing it with one using the latest processor technology.

The ability to run hundreds of laboratory experiments in parallel with little or no human intervention is now possible with modular robotic components, standardized controllers and computer-vision-enabled monitoring and data collection.  Of course, prior to automating a complicated endeavor like that addressed by the Encyclopedia of DNA Elements (ENCODE) Consortium\footnote{%
  The goal of ENCODE is to build a comprehensive parts list of functional elements in the human genome, including elements that act at the protein and RNA levels, and regulatory elements that control cells and circumstances in which a gene is active.}, you first have to do a lot of exploratory work to figure out what's worth automating.
That said, researchers should be aware of the potential advantages of automation tools and quick to exploit them when they identify an appropriate task.  The short- and medium-term prospects for more parallelism, higher throughput, greater precision and enhanced flexibility is limited primarily by our imagination and willingness to invest in building and deploying the requisite systems.

% %%%%%%%%%%%%%%%%%%%%%%%%%%%%%%%%%%%%%%%%%%%%%%%%%%%%%%%%%%%%%%%%%%%%%%%%%%%%%%%%%%%%%%%%%%%%%%%%%%%%%%%%%%%% %

%% file: sequencing.tex
\section{Synthetic Neurobiology}
\label{section:sequencing}

% %%%%%%%%%%%%%%%%%%%%%%%%%%%%%%%%%%%%%%%%%%%%%%%%%%%%%%%%%%%%%%%%%%%%%%%%%%%%%%%%%%%%%%%%%%%%%%%%%%%%%%%%%%%% %

% http://en.wikipedia.org/wiki/Conserved_sequence

In this section, we consider the methodology of co-opting existing biomolecules for purposes other than those for which they naturally evolved.  This approach has the advantage that it often simplifies the problem of biocompatability.  Moreover it allows us to take advantage of the enormous diversity of solutions to biologically relevant problems provided by natural selection~\cite{RuschetalPLoS-07}.  Given that cellular function is conserved across a range of organisms from algae to mice, if an existing solution from the target organism is not found, a solution from an alternative organism is often compatible.

% http://en.wikipedia.org/wiki/Fluorophores

We've already seen several examples of biomolecules that play important roles in technologies relevant to scalable neuroscience.  For instance, organic fluorophores are used in imaging technologies to stain tissues or as markers for active reagents, as in the case of antibodies used in immunoflorescence imaging.  Biomolecules found in odd flora and fauna have found application in genomics and provide a dramatic example of how biology can enable exponential scaling.

% http://en.wikipedia.org/wiki/Taq_polymerase

A key step in DNA sequencing involves amplifying the DNA to create the many copies required in subsequent steps such gel electrophoresis\footnote{%
  It is worth noting that this amplification step will no longer be necessary in nano­pore sequencing or other single­molecule sequencing methods~\cite{VentaetalNANO-13}.}.
Amplification requires multiple cycles in which a heat-stable DNA polymerase plays a critical role.  An important breakthrough was the discovery of Taq polymerase isolated from {\it{Thermus aquaticus}} a species of bacterium that can tolerate high temperatures, but it required considerable additional effort before the method now called polymerase chain reaction (PCR) was refined and became an indispensable tool in genomics~\cite{MullisSCIAM-90}.

% http://en.wikipedia.org/wiki/Channelrhodopsin

The discovery of complex molecules called {\it{channelrhodopsins}} played a similar role in spurring the development of optogenetics~\cite{BoydenetalNATURE-05}.  These molecules function as light-gated ion channels and serve as sensory photoreceptors in unicellular green algae to control movements in response to light. We'll look at the development of optogenetics in more detail as a case study in how technologies in this field develop and a object lesson in the difficulty of predicting how long it takes to propel an idea from its initial conception to a useful technology. 

% %%%%%%%%%%%%%%%%%%%%%%%%%%%%%%%%%%%%%%%%%%%%%%%%%%%%%%%%%%%%%%%%%%%%%%%%%%%%%%%%%%%%%%%%%%%%%%%%%%%%%%%%%%%% %

Once you've identified a suitable molecule to perform your target function, you have to figure out how to introduce it into the cell. In some cases, it is possible to introduce the molecule directly into the intra- or extra-cellular fluid, but more often than not, it is necessary to enlist the cell's machinery for replication, transcription and translation to produce the necessary active molecules and control their application.  This last typically involves the technology of recombinant DNA and synthetic biology.

There are a number of technologies for inserting genes into a host cell's DNA.  In some cases, you can co-opt an existing pathway that will express and control the expression of a protein in exactly the right way.  It is also possible to modify an existing pathway or create an entirely new pathway, but such modifications and additions are notoriously difficult to get right and can delay or derail development.  Such problems are exacerbated in the case of introducing additional compounds into the mix. 

Despite its challenges, this approach to building nanoscale recording and reporting devices is promising since the issues of biocompatibilty which plague more exotic approaches based on nanotechnology are much more readily solved, given the current state of the art.  Moreover, these solutions utilize existing cellular sources of energy and biomolecular processes, many of which can be viewed as carrying out information-processing tasks, are remarkably energy efficient compared to semiconductor technology.  To illustrate, we look at three examples:

% %%%%%%%%%%%%%%%%%%%%%%%%%%%%%%%%%%%%%%%%%%%%%%%%%%%%%%%%%%%%%%%%%%%%%%%%%%%%%%%%%%%%%%%%%%%%%%%%%%%%%%%%%%%% %

The role of competition, rich collaboration, timing, serendipity and just plain luck make it difficult to predict when or even whether an idea, however compelling, will mature to point that it serves as a useful tool to enable new science. This is nowhere more apparent than in the case of optogenetics, one of the most powerful new technologies to emerge out of systems neuroscience in the last decade~\cite{YizharetalNEURON-11}.  

The basic molecules used in optogenetics --- called {\it{opsins}} --- have been studied since the 1970s.  These complex proteins undergo conformational changes when illuminated that serve to control the transfer of specific ions across the membranes of cells in which they are expressed.  Found in archaea, bacteria, fungi and algae these large-molecule proteins serve diverse photosynthetic and signaling purposes.  While obvious in hindsight, at the time it was not obvious that these proteins could be re-purposed to control the firing of neurons. 

Once the basic idea was formulated, the search was on for a molecule that could be expressed in mammals at levels high enough to mediate neural depolarization, was well tolerated at such levels, didn't require additional reagents to enable {\it{in vivo}} experiments, and recovered quickly enough to allow for precise control of the target neurons.  There were also the technical problems of how to introduce the molecule into cells and how to precisely deliver light to the target neurons that needed to be solved to provide a compelling demonstration.

Before the field stumbled upon channelrhodopsin there were a bunch of alternative technologies being considered including ones with longer chains of dependencies and more complicated component technologies involving both natural and synthetic options both biological and inorganic.  This is often the natural chain of events leading to a simpler, more elegant solution matching the task to a specific solution in nature that solves the problem. 

Once the idea was out, several labs came out with related technologies around the same time. There were also plenty of exciting innovations including a demonstration of light-activated neural silencing in mammals and refinements such as a channelrhodopsin that can be opened by a blue light pulse and closed by a green or yellow light pulse.  Opsins probably could have been applied in neuroscience decades earlier~\cite{BoydenetalNATURE-05}, but excitement and competition often serve as the tinder required to ignite a flurry of rapid development~\cite{ZhangetalNATURE-10}. 

% %%%%%%%%%%%%%%%%%%%%%%%%%%%%%%%%%%%%%%%%%%%%%%%%%%%%%%%%%%%%%%%%%%%%%%%%%%%%%%%%%%%%%%%%%%%%%%%%%%%%%%%%%%%% %

Generating the connectome at the macroscale --- tracing the major white-matter connections between functional areas --- is reasonably tractable using diffusion tensor MRI.  Generating the microscale connectome --- mapping the connections between individual neurons --- is much more challenging and it may be that approaches based on analyzing scanning electron micrographs are tackling a harder problem than is strictly necessary.  Might there be a simpler way of creating a catalog of all the neurons in a tissue sample, their connections, cell type, connection strengths and 3-D coordinates?

Zador~\etal~\cite{ZadoretalPLoS-12} have proposed the idea of ``sequencing the connectome'' in a paper of the same title emphasizing the opportunity to leverage one of the most important scalable technologies to emerge from the biological sciences.  The basic concept is simple.  The authors break down the problem into three components: (a) label each neuron with a unique DNA sequence or ``barcode''', (b) propagate the barcodes from each source neuron to each synaptically-adjacent sink neuron --- this results in each neuron collecting a ``bag of barcodes'', and (c) for each neuron combine its barcodes in source-sink pairs for subsequent high-throughput sequencing.  

All three steps --- excluding the sequencing --- are to be carried in the live cells of the target tissue sample.  The details are not nearly as simple and each step will likely require significant engineering effort.  However, as in the case of optogenetics, there are a number of tools that might be applied directly or adapted to suit.  There exist reasonable approaches for generating random barcodes from DNA templates that could be used as unique designators for individual neurons.  

We also know how to propagate barcodes transynaptically using viruses that naturally spread from neuron to neuron through synapses. The operation of ``concatenating'' barcodes is carried out in any number of neurons. Sequencing and location tagging would be accomplished using the method described in Section~\ref{section:probing} for generating a static volumetric transcriptomic map. Getting all these pieces to play together and not compromise the cell before all the connectomic data is collected might be the work of a summer or several years.  It is difficult to predict the outcome, but several experts believe Zador's approach or something like it will work and the first demonstrations on simple organisms are one to two year out.

% %%%%%%%%%%%%%%%%%%%%%%%%%%%%%%%%%%%%%%%%%%%%%%%%%%%%%%%%%%%%%%%%%%%%%%%%%%%%%%%%%%%%%%%%%%%%%%%%%%%%%%%%%%%% %

Calcium imaging offers our best near-term scalable method for recording from large ensembles of living neurons in mice or even simpler experimental animals.  But what if we want to record from neurons deep within the brain in an awake behaving human?  One proposal is to design a molecular machine to sense a convenient correlate of neural activity such as elevated calcium concentrations and write the resulting data to a molecular ``ticker tape''~\cite{ZamftetalPLoS-12,KordingPLoS-11}.  While somewhat more subtle than the previous proposal, the individual steps are conceptually straightforward.

Cellular transcoding processes such as DNA replication and messenger RNA transcription make occasional errors in the process of incorporating nucleotides into their respective products. The proposed method depends on harnessing an enzyme called DNA polymerase (DNAP) that plays a central role in DNA replication. Relying on the property that the rate of misincorporation is dependent on the concentration of positive ions or {\it{cations}}. They show that by modulating cation concentrations one can influence the misincorporation rate on a reference template in a reliable manner so that information can be encoded in the product of DNAP and then subsequently recovered by sequencing and comparing with the reference template.

There are a lot of moving parts in the ticker-tape technology. A full solution will require packaging all the necessary enzymes, nucleotide-recycling, and metabolic machinery necessary to initiate and sustain DNAP reactions outside of the nucleus.  This idea may be in the early, wishful-thinking stage of exploring complex, multi-step solutions, hoping that in the meantime some molecular biologist will stumble across an existing ticker-tape-like solution just waiting for us to exploit\footnote{%
  Finding a DNAP sensitive to calcium or other biologically relevant ions is difficult, but, as in the case of channelrhodopsin and Taq polymerase, a novel organism's DNAP might come to the rescue, especially as it becomes clearer exactly what we're looking for.}.  
It is very hard to make any reasonable predictions of how long it will take this technology to mature or if it will resemble anything like the current proposal if a solution is ever realized.  Two to five years is optimistic and a final solution will likely bear little relationship to the current proposal. 

% %%%%%%%%%%%%%%%%%%%%%%%%%%%%%%%%%%%%%%%%%%%%%%%%%%%%%%%%%%%%%%%%%%%%%%%%%%%%%%%%%%%%%%%%%%%%%%%%%%%%%%%%%%%% %

It is no longer necessary to depend solely on what we discover in the natural world when searching for biomolecules for engineering purposes.  Methods from synthetic biology like rational protein design and directed evolution~\cite{BornscheuerCOCB-01} have demonstrated their effectiveness in synthesizing optimized calcium indicators for neural imaging~\cite{AkerboometalNEUROSCIENCE-12} starting from natural molecules. Biology provides one dimension of exponential scaling and additional performance can be had by applying high-throughput screening methods and automating the process of cloning and testing candidates solutions.  

There are also opportunities for accelerated returns from improved algorithms for simulating protein dynamics to implement fast, accurate, energy function used to distinguish optimal molecules from similar suboptimal ones. Advances in this area undermine arguments that quantum-dots and related engineered nanoscale components are crucial to progress because these technologies can be precisely tuned to our purposes.  Especially when weighed against the challenges of overcoming the toxicity and high-energy cost of these otherwise desirable industrial technologies.

% %%%%%%%%%%%%%%%%%%%%%%%%%%%%%%%%%%%%%%%%%%%%%%%%%%%%%%%%%%%%%%%%%%%%%%%%%%%%%%%%%%%%%%%%%%%%%%%%%%%%%%%%%%%% %

Biology doesn't offer any off-the-shelf solutions addressing the reporting problem aside from writing the recorded data to some stable molecular substrate like DNA and then flushing the encapsulated data out of the nervous system to be subsequently recovered from the organism's blood or urine\footnote{%
  Specifically, we know of no practical method to induce exocytosis of large amounts of DNA that doesn't involve killing the cell.}
We aren't likely to find solutions to the problem of transmitting large amounts of data through centimeters of nervous tissue for the obvious reason that nature hasn't come across any compelling need to read out the state of large neural populations.  To engineer scalable solutions to the reporting problem, we may have to look to advances in building nanoscale electronic circuits. 

% %%%%%%%%%%%%%%%%%%%%%%%%%%%%%%%%%%%%%%%%%%%%%%%%%%%%%%%%%%%%%%%%%%%%%%%%%%%%%%%%%%%%%%%%%%%%%%%%%%%%%%%%%%%% %

%% file: shrinking.tex
\section{Nanotechnology}
\label{section:shrinking}

% %%%%%%%%%%%%%%%%%%%%%%%%%%%%%%%%%%%%%%%%%%%%%%%%%%%%%%%%%%%%%%%%%%%%%%%%%%%%%%%%%%%%%%%%%%%%%%%%%%%%%%%%%%%% %

Nanotechnology, the manipulation of matter at the atomic scale to build molecular machines, is already contributing to brain science~\cite{AlivisatosetalNANO-13}.  Much of the recent progress in developing probes for multi-cell recording depends on nanotechnology~\cite{ZorzosetalOPTICS-12,BarrettoandSchnitzerCSHP-12,DueetalPLoS-11,ChenetalMEMS-10}.  From advances in chip fabrication for faster computers to better contrast agents for imaging and high-density, small-molecule assays for high-throughput pathogen testing, the products of nanotechnology research play an increasingly important role in the lab and clinic~\cite{ZhangandWangTHERANOSTICS-12,WangetalJBO-12,WuetalNANO-08,CaietalNANOSCALE-07}.  

It isn't so much a matter of {\it{whether}} nanotechnology will deliver implantable whole-brain scanning, but {\it{when}}, and, unfortunately, predicting when is complicated.  On the one hand, it is possible to make some rough predictions about when we can expect a chip with a transistor count of 10,000, peak power consumption less than 5 nanowatts and size small enough to enter the brain through the capillaries might be feasible --- anywhere from five to ten years depending on the degree to which Moore's law can be sustained\footnote{%
  Here we are assuming that power consumption per device will continue to fall by half every 1.5 years, at least for the next five or so years.  Koomey's law~\cite{KoomeyetalIEEE-11} describing this trend was based on performance over the last six decades but there are some worrisome issues projecting forward, especially when one considers the energy consumed in charging and discharging capacitance in transistors and interconnect wires and the static energy of devices in terms of controlling leakage current~\cite{TuckerandHintonIEEE-11}. If progress is stalled or seriously retarded, the prospects for keeping to our ambitious time line will obviously be negatively impacted.}.

On the other hand, issues of toxicity and heat dissipation pose problems likely to delay approval for use in humans, not to mention the consequences of implanting millions of computer chips in your brain.  In this section, we look at what is possible now and what we believe will be possible in the medium- to longer-term horizon.  As a thought experiment, we sketch a not-entirely-implausible technology for whole-brain scanning to focus attention on some of the technological challenges.  Think of a brain-scale cellular network with recorders playing the role of mobile phones, local reporters that of cell towers, and relays that of phone-company switches or networked computers.~\footnote{%
  It may one day be feasible to install a network of nanoscale wires and routers in the extracellular matrix. Such a network might even be piggybacked on the existing network of fibrous proteins that serve as structural support for cells, in the same way that the cable companies piggyback on existing phone and power distribution infrastructure.  At one point, we discussed the idea of using kinesin-like ``wire stringers'' that would walk along structural supports in the extracellular matrix to wire the brain and install a fiber-optic network.}
The purpose of this exercise is to explore the boundary between science fact and science fiction in order to better understand the promise and peril inherent in this rapidly evolving technology. 

% %%%%%%%%%%%%%%%%%%%%%%%%%%%%%%%%%%%%%%%%%%%%%%%%%%%%%%%%%%%%%%%%%%%%%%%%%%%%%%%%%%%%%%%%%%%%%%%%%%%%%%%%%%%% %

The technology we envision requires several types of implantable devices operating at different scales.  At the smallest scale, we imagine individual cells instrumented with recorder molecules that sense voltage levels, proteins, expression levels, and perhaps changes in cell structure.  There already exist micro-scale MEMS {\it{lab-on-a-chip}} devices employing scalable technologies that should serve to produce nanoscale variants as the fabrication techniques mature~\cite{ZhangetalSENSORS-13,KorlachetalPNAS-08,MannionandCraigheadBIOPOLYMERS-07,CraigheadNATURE-06}. These recording devices would transmit signals to be read by nanoscale reporting devices floating in the surrounding extracellular fluid or anchored at strategic locations in the extracellular matrix.

Imagine one type of reporting device distributed throughout the brain so there is one such device within a couple of microns of every neuron.  Each such reporter is responsible for picking up the signals broadcast from recorders in its immediate vicinity.  These reporting devices might directly transmit information to receivers located outside the brain but we anticipate this approach running into power and transmission limitations.  Instead, suppose that these local reporters participate in a heterogeneous network of devices allowing for different modes of transmission at different scales.

% http://en.wikipedia.org/wiki/Dura_mater, http://en.wikipedia.org/wiki/Ventricles_of_brain

The local reporters would forward information to a second type of reporting device more sparsely distributed and responsible for relaying the information received from the local reporters to external receivers not limited by size or power requirements.  These relay devices could number in the thousands instead of millions, be somewhat larger than the more numerous and densely distributed reporters, and be located inside of the dura but on the surface of the brain, within fissures or anchored on the membranes lining the capillaries supplying blood to the brain and the ventricles containing cerebrospinal fluid.

% http://www.technologyreview.com/view/421153/how-close-is-a-workable-brain-computer-interface/

Recorders might use optical signalling or even diffusion to transmit information to local reporting devices.  The local reporters might employ limited-range photonic or near-field technologies to forward information to the nearest relay device.  Finally, the relays multiplex information from many reporters and forward the result using microscale RF transmitters or micron-sized optical fibers connected to a ``Matrix''-style physical coupling.  

Relays located on the surface of the brain would have more options for power distribution and heat dissipation, and, being larger than local reporters, they would stand a better chance of having enough room to integrate efficient antennas. The RF transmitters would have no problem with scattering and signal penetration, and, because there are orders-of-magnitude fewer relays than local reporters, it is more likely that the available frequency spectrum will be able to supply sufficient bandwidth. Two-way transmission would enable sensing and the ability to excite or inhibit individual neurons.

% http://en.wikipedia.org/wiki/BrainGate#Research_and_Experimental_Results

Ignoring the potential health risks associated with such an invasive technology, most of the components mentioned above only exist in science fiction stories.  That said, something like the above scenario is well within the realm of possibility in the next twenty years.  Moreover, one can easily imagine less-invasive versions for medical applications in which the benefits outweigh the risks to patients.  Indeed, we are already seeing clinical trials for related technological advances that promise to restore mobility to patients with severe disabilities~\cite{BansaletalNEUROPHYSIOLOGY-12,GiljaetalNATURE-12,TruccoloetalNEUROSCIENCE-08}. 

% %%%%%%%%%%%%%%%%%%%%%%%%%%%%%%%%%%%%%%%%%%%%%%%%%%%%%%%%%%%%%%%%%%%%%%%%%%%%%%%%%%%%%%%%%%%%%%%%%%%%%%%%%%%% %

Let's consider each of the components in our {\it{Gedankenexperiment}}, focusing on the reporting problems. The recorders associated with individual cells might encode data in packets shunted into the extracellular fluid for transmission to local reporters by diffusion.  Suppose the packets are the size (\textasciitilde{}50 nm) and diffusivity (\textasciitilde{}10$^{-8}$ cm$^{2}$/s) of vesicles used for cellular transport and suppose there are a million local reporters distributed uniformly throughout a human brain.  It would take about a day on average to transport a packet from a recorder to a nearby reporter~\cite{Sheets2008Biophys}. 

A biologically-motivated alternative to decrease transit time and reduce the number of reporters might involve equipping each vesicle with a molecular motor or {\it{flagella}} providing an effective diffusivity of around 10$^{-5}$ cm$^{2}$/s~\cite{Bush2010,WeibeletalPNAS-05}.   With only 1000 reporters evenly distributed throughout the brain, it would take on average less than 2 hours for these augmented vesicles to traverse the necessary distance. Packing enough information into each vesicle in the recorder and unpacking, decoding and relaying the information in the reporter fast enough to support the necessary bandwidth is quite another challenge, but leveraging the molecular machinery of DNA replication and transcription might offer a possible solution  --- see Appendix~\ref{appendix:daniel}.

Each reporter would transmit this information to the closest relay device on the surface of the brain.  Depending on the application, we might want to distribute more relay devices on the surface of the cortex and thalamus than elsewhere.  The surface area of an adult brain is around 2,500 cm$^2$.  Given a thousand relays, each relay is responsible for a couple of square centimeters. Assuming a million local reporters proportionally distributed, each relay is responsible for forwarding information from a thousand local reporters.  We would need each reporter to transmit up to $2.5 \times \mbox{\rm{sqrt}}(2)/2$ cm = 1.76 cm. 

Unfortunately, we can't position the reporters farther than about 500 $\mu$m and still retain the benefits of optical frequencies: low absorption to minimize tissue damage, high frequency to optimize bandwidth.  Pushing aside these challenges, suppose there exists a frequency band in which scattering and absorption are low in brain tissue and relay devices along with their associated antennae are small enough to allow implantation. 

If we allow for 10 bits per millisecond per neuron on average assuming only a fraction of neurons will have anything to report at any given time, 50\% overhead for packetization, addressing, error correction, etc, then we arrive at 15 Gb per second as a bound on the channel capacity required of each relay. In any given millisecond, about 1\% or 10,000 of the million reporters will be communicating with their respective relay --- assume that neurons fire at 10 Hz and so only 10 milliseconds of every 1000 actually have a firing peak and require a signal. 

% http://en.wikipedia.org/wiki/Orthogonal_frequency-division_multiplexing

For high-enough carrier frequencies, suppose we can use a modulation scheme like orthogonal-frequency-division-multiplexing (OFDM)~\cite{Le2012}, splitting up a congested bandwidth into many narrow-band signals. Then, assuming a reasonably powerful and compact source of Terahertz radiation, we could again apply OFDM and give each relay its own 15 GHz of bandwidth without inhibiting transmission --- see Appendix~\ref{appendix:andre}. 

As an alternative to relying entirely on the electromagnetic (EM) spectrum, researchers have proposed an approach in which the sub-dural reporters communicate with and supply power to a class of (implanted) reporters using ultrasonic energy~\cite{SeoetalARXIV-13}.  
The implanted reporters are referred to as {\it{neural dust}} and we'll refer to an individual implanted reporter as a {\it{dust mote}}.  In this proposal, dust motes are also responsible for recording measurements of the local field potential (LFP) that the dust motes then transmit to the sub-dural reporters.  The dust motes are approximately cube shaped, measuring 50~$\mu$m on a side, and consist of a piezoelectric device and some electronics for powering the mote and transmitting LFP measurements.  

A 100~$\mu$m mote designed to operate at a frequency of 10~MHz, $\lambda = 150$~$\mu$m has~\textasciitilde{}1~dB attenuation at a 2 mm depth --- the attenuation of ultrasound in neural tissue is~\textasciitilde{}0.5 dB/(cm MHz).  A comparable electromagnetic solution\footnote{%
  Note that the speed of sound --- approximately 500 m/s in air and 1500 m/s in water --- is considerably slower than that of an electromagnetic signal --- exactly 299,792,458 m/s in a vacuum and close enough to that for our purposes in most other media. The relatively slow acoustic velocity of ultrasound results in a substantially reduced wavelength when compared to an electromagnetic signal at the same frequency.  Compare for example a 10~MHz, $\lambda = 150$~$\mu$m ultrasound signal in water with a 10 Mhz, $\lambda = 30$m EM signal. An EM signal of this wavelength would be useless for neural imaging; the tissue would be essentially transparent to the signal and so penetration depth would be practically unlimited, but there would hardly be any reflected signal and the size of an antenna necessary to receive such a signal would be prohibitively large --- on the order of half the wavelength for an efficient antenna.  A comparable EM solution for neural dust~\cite{SeoetalARXIV-13} would therefore be closer to the 10~GHz frequency provided in the text.} 
operating at 10~GHz, $\lambda = 5$~mm has~\textasciitilde{}20 dB attenuation at 2 mm.  There are other complicating factors concerning the size of the inductors required to power EM devices and efficient antennae for signal transmission --- displacement currents in tissue and scattering losses can increase attenuation to 40 dB in practical devices.  The proposed approach posits that electrophysiological data might be reported back via backscattering by modulating the reflection of the incident carrier wave used to supply power. 

The small size of the implanted devices also complicates directly sensing the local field potential.  The electrodes employed by electrophysiologists measure local field potentials at one or more locations with respect to a second, common electrode which acts as a ground and is located at some distance from the probe.  In the case of LFP measurements made by dust motes, the distance between the two electrodes is constrained by the size of the device such that, the smaller this distance, the lower the signal-to-noise ratio --- see Gold~\etal~\cite{GoldetalJCN-07} for an analysis of extra- and intra-cellular LFP recording strategies for constraining compartmental models.  The S/N problem can be ameliorated somewhat by adding a ``tail'' to each dust mote thereby locating the second electrode at some distance from the first one situated on the main body.

% http://en.wikipedia.org/wiki/Terahertz_radiation

It is hard not to be a bit cavalier in fending off the technical challenges faced in trying to read off anything approaching the full state of an active brain, but the main purpose of our thought experiment is to acquaint the reader with some of these challenges, convey a measure of just how audaciously hard the problem is, and give some idea of the wide range of techniques drawn from many disciplines that are being brought to bear in attempting to solve it~\cite{MarblestoneetalARXIV-13}. 

% %%%%%%%%%%%%%%%%%%%%%%%%%%%%%%%%%%%%%%%%%%%%%%%%%%%%%%%%%%%%%%%%%%%%%%%%%%%%%%%%%%%%%%%%%%%%%%%%%%%%%%%%%%%% %

% http://en.wikipedia.org/wiki/Lentivirus}{lentivirus} 

Noting that any of the nanotechnologies discussed would require some method of ``installation'' and assuming that our encouraging the development of ``designer babies'' with pre-installed recording and reporting technologies would be awkward and likely misunderstood, we came up with the following simple protocol for installing a generic nanotechnology solution along the lines of our {\it{Gedankenexperiment}}:
\begin{enumerate}
\item modification of individual cells to express biomolecular recorders might be accomplished using a lentivirus --- unique among retroviruses for being able to infect non-dividing cells --- or alternative viral vector and recombinant-DNA-engineered gene payload; while not a nanotechnology solution {\it{per se}}, this sort of biological approach will likely remain the best option for instrumenting individual neurons for some time to come, 
\item distribution of local reporting devices might be accomplished using the circulatory system and, in particular, the capillaries\footnote{%
    Needless to say it will be tricky shrinking the reporter technology to a size small enough that it can pass easily through the capillaries without inducing a stroke.} 
in the brain as a supply network with steps taken to alter the blood-brain-barrier using focused ultrasound~\cite{HynynenetalNEUROIMAGE-5l} during the installation process; once in the brain the reporters could be anchored in the extracellular matrix using techniques drawn from {\it{click chemistry}}~\cite{KolbetalCLICK-01},
\item pairing of neurons with their associated reporter devices might be accomplished using IPTG (isopropyl $\beta$-D-1-thiogalactopyranoside), tamoxifen, or other methods of controlling gene expression to achieve pairing; the associated machinery could be bundled with the lentivirus payload or introduced using a separate helper virus; this sort of {\it{in situ}} interface coupling between biological and nanotech components is likely to become common in hybrid solutions, and, finally, 
\item installing a grid of relay devices on the surface of the brain without a craniotomy is likely to remain challenging until such time as nanotechnology develops nanorobotic devices capable of navigating in tissue either autonomously or via some form of teleoperation; in the meantime, there may be applications in which such a grid could be installed, at least on the cortical surface employing the same procedure used for intraoperative electrocorticography~\cite{SchuhandDrurySAPMP-97}.
\end{enumerate}
It is worth noting that a more natural method of ``installation'' might indeed involve a developmental approach patterned after the self-assembly processes governing embryonic development, but so far our understanding of molecular self assembly is limited to the simplest of materials such as soap films and, even in these cases, nature continues to surpass our best efforts.

% %%%%%%%%%%%%%%%%%%%%%%%%%%%%%%%%%%%%%%%%%%%%%%%%%%%%%%%%%%%%%%%%%%%%%%%%%%%%%%%%%%%%%%%%%%%%%%%%%%%%%%%%%%%% %

The development of technology for building nanoscale communication networks~\cite{Bush2010} could propel our fictional brain-computer-interface into the realm of the possible.  We're already seeing prototypes for some of the most basic components required to build communication networks, including carbon nanotubes to construct RF~\cite{KocabasPNAS-08} and optical antennas~\cite{KempaetalAM-07} and multiplexers fabricated from piezoelectric nanomechanical resonators that could enable thousands of nanoscale biosensors to share the same communication channel~\cite{SadeketalNANO-10}.  It is possible to build a fully-functional radio receiver from a single carbon nanotube~\cite{JensenetalNANO-07} and the possibilities for optical communication employing more exotic technologies from the field of plasmonics~\cite{SchulleretalNATURE-10} are even more intriguing.  However, as promising as these demonstrations may seem, practical developments in this arena appear to be some years off. 

% %%%%%%%%%%%%%%%%%%%%%%%%%%%%%%%%%%%%%%%%%%%%%%%%%%%%%%%%%%%%%%%%%%%%%%%%%%%%%%%%%%%%%%%%%%%%%%%%%%%%%%%%%%%% %

One area where nanotechnology is likely to have a big impact is in computing hardware.  Logic circuits and non-volatile memory fabricated from carbon nanotubes are among our best hopes for sustaining Moore's law beyond the next few years.  The prospects are excellent for smaller, faster, lower-power devices based on carbon nanotubes with possible spin-off benefits for neuroscience.  Carbon nanotubes can be conjugated with DNA to render them biocompatible and it may be possible to embed these composites in cell membranes to be used as voltage sensors and to excite and inhibit individual neurons, thereby providing a practical alternative to silicon and other semiconductor materials that are toxic to cells. 

% %%%%%%%%%%%%%%%%%%%%%%%%%%%%%%%%%%%%%%%%%%%%%%%%%%%%%%%%%%%%%%%%%%%%%%%%%%%%%%%%%%%%%%%%%%%%%%%%%%%%%%%%%%%% %

%% file: investing.tex
% \rawhtml
% <a name="technology_investment_opportunity"></a>
% \endrawhtml
% \section{Technology Investment}
\label{section:investing}

Commercial opportunities and industry involvement related to a scientific endeavor can help to drive the development of enabling technologies and accelerate progress.  In the case of the Human Genome Project, commercial interest lagged scientific progress in the early days, but once the business opportunities became apparent industry, involvement accelerated with collateral benefits for science in the form of better, cheaper and faster sequencing and related tools and techniques~\cite{Davies2010,ChurchSCIAM-06}. In this section, we briefly survey some of the opportunities that might drive investment and spur innovation in neuroscience.

Companies that offer products and services relating to scalable neuroscience abound. 
%
% http://www.3scan.com/, http://www.gatan.com/, http://www.inscopix.com/
%
There are companies like FEI, Gatan, Inscopix, 3Scan and Zeiss that specialize in electron microscopy, stains and preparations, and automation for serial-sectioning. 
%
% http://neuralynx.com, http://genetherapy.unc.edu/services.htm
%
Those that primarily serve the research community like Neuralynx and UNC Joint Vector Laboratories by providing electrophysiology tools for research including hardware and reagents for optogenetics.
%
% http://www.nanoimmunotech.eu/en/
%
Industrial products and services like those offered by Nanoimmunotech including off-the-shelf and special-order nanoparticles and technologies for joining (conjugating) nanostructures, dyes, biomolecules.  
%
% http://www.openoptogenetics.org/,  http://openeeg.sourceforge.net/
%
There are even grass-roots open-source communities like Open Optogenetics and Open EEG that create communities and offer educational resources and open-source tools and protocols. 
And then there is the incredible array of companies large and small that provide reagents, cell lines, hardware, software and services to the medical community. 

Medicine has long been a driver for technology with ample capital funding to underwrite the development costs of new technologies.  Assays for analyzing diseased tissue, tumors, cell counts, DNA sequences all have their neurophysiological counterparts, and, in the case of the brain, there is a felt need for lower-cost, non-invasive, readily-accessible diagnostic tools.  The incumbent technology providers will likely maintain their technical and market advantages, but big companies tend to be slow to innovate and generally fail to offer incentives to their engineers to take on high-risk, high-payoff projects. Bulky, expensive hardware like MEG, MRI, PET, high-end EEG and ultrasound imaging and FUS equipment offer significant competitive challenges to the small company.  That said there are opportunities in the synthetic biology and nanotechnology arenas to transition research ideas to products that could, in time, challenge even these markets. 

% %%%%%%%%%%%%%%%%%%%%%%%%%%%%%%%%%%%%%%%%%%%%%%%%%%%%%%%%%%%%%%%%%%%%%%%%%%%%%%%%%%%%%%%%%%%%%%%%%%%%%%%%%%%% %

% http://www.myzeo.com, http://www.emotiv.com, http://neurosky.com

In the near-term, companies like Neurosky, Emotiv, InterAxon, Zeo and Hitachi are pursuing BCI opportunities that leverage inexpensive, portable, non-invasive, off-the-shelf technologies such single-chip multi-channel EEG, EMG, NIRS, eye- and gaze-tracking, microfluidic immunoassay chips, etc. to provide tools and consumer devices for meditation, entertainment, sleep management, and out-patient monitoring.  Near-infrared spectroscopy (NIRS) is a good example of a relatively inexpensive, non-invasive technology for measuring functional hemodynamic responses to infer neural activation --- correlated with fMRI BOLD signals --- which could be integrated with wearable displays like Google Glass.  

One target of particular consumer interest is the development of personal assistants that know us better than we do, can help us to calibrate our preferences and overcome instinctive biases to make better decisions, and allow us to monitor or even exercise some degree of control over our emotional states to overcome anxiety or depression.  Building such intimate software assistants and accompanying sensor suites that are comfortable and even fashionable is challenging. Moreover a successful product will have to cope with the fact that the class of sensors that are practical for such applications will offer only rough proxies for those emotional and physiological markers likely to be most useful in understanding our moods and preferences.

% %%%%%%%%%%%%%%%%%%%%%%%%%%%%%%%%%%%%%%%%%%%%%%%%%%%%%%%%%%%%%%%%%%%%%%%%%%%%%%%%%%%%%%%%%%%%%%%%%%%%%%%%%%%% %

% http://stanford.edu/group/scsnl/cgi-bin/drupal_scsnl/

There is reason to believe that existing options for sensing can be combined using machine learning to predict the relevant psychophysical markers.  The obvious approach would be to create a training set using more expensive technologies such as MRI, MEG, and higher-quality, multi-channel EEG.  Signatures for healthy and pathological variants of many cognitive functions are now well established to the point where we can predict common brain states by analyzing fMRI images.  Researchers have demonstrated inter-subject synchronization of fMRI responses during natural music listening, providing a baseline that could be used to predict musical preferences\cite{AbramsetalEJN-13}.  The development of activity based classifiers to identify functional areas in animal models~\cite{BosmanetalNEURON-12} could could be extended to take ECoG or EEG data from humans and do the same.

There are opportunities for small startups savvy in user-interface and machine-learning technologies to partner with neurophysiologists and cognitive neuroscientists to build products for personalized medicine, assistive technology and entertainment.  Researchers at New York University and Johns Hopkins have shown that concussion, dementia, schizophrenia, amyotrophic lateral sclerosis (ALS), autism and Fragile X syndrome (FXS) are among the numerous diseases with characteristic anomalies detectable using eye movement tracking technology~\cite{Newman-TokerSTROKE-13,SamadaniandHeeger2013}. Researchers at Yale used fMRI~\cite{ScheinostTP-13} to enable subjects to control their anxiety through a process of trial and error resulting in changes that were still present several days after the training. The experimental protocol relied on displaying the activity of the orbitofrontal cortex (a brain region just above the eyes) to subjects while they lay in a brain scanner.

% %%%%%%%%%%%%%%%%%%%%%%%%%%%%%%%%%%%%%%%%%%%%%%%%%%%%%%%%%%%%%%%%%%%%%%%%%%%%%%%%%%%%%%%%%%%%%%%%%%%%%%%%%%%% %

% http://en.wikipedia.org/wiki/Transcranial_magnetic_stimulation

In the medium-term, there will be opportunities using transcranial magnetic stimulation, focused ultrasound, implantable electrical and optogenetic arrays for patients with treatment-resistant anxiety, depression, stroke, head-injury and tumor-related tissue damage, neourodengenerative diseases, etc.  Here again partnering with researchers, sharing IP with academic institutions, and finding chief executives and venture capital partners experienced with health-related technologies will be key.   Longer-term opportunities for better prosthetics, cognitive and physical augmentation, entertainment, etc.\ are the stuff of science fiction but could arrive sooner than expected if key technologies mature more quickly than anticipated. 

% %%%%%%%%%%%%%%%%%%%%%%%%%%%%%%%%%%%%%%%%%%%%%%%%%%%%%%%%%%%%%%%%%%%%%%%%%%%%%%%%%%%%%%%%%%%%%%%%%%%%%%%%%%%% %

As for the considerable promise of nanotechnology, venture capital firms interested in this area might want to hedge their bets by dividing investment between (1) non-biological applications in communications and computing where biocompatibility isn't an issue and current technologies are up against fundamental limitations, and (2) biological applications in which the ability to design nanosensors with precisely-controllable characteristics is important.

Regarding (1) think in terms of quantum-dot (QD) lasers, photonics, and more-exotic-entangled-photon technologies for on-chip communication --- 2-D and 3-D chips equipped with energy-efficient, high-speed interprocessor communication supporting dense core layouts communicating using arbitrary, even programmable topologies.  Regarding (2) there is plenty of room for QD alternatives to natural chromophores in immunoflorescence imaging~\cite{MazumderetalJoN-09}, voltage-sensing recording~\cite{MarshallandSchnitzerACS-13}, and new contrast agents for MRI~\cite{ParatalaetalPLoS-12}.  Advances in precisely controlling QD properties will help to fuel the search for better methods of achieving biocompatibility. 

% %%%%%%%%%%%%%%%%%%%%%%%%%%%%%%%%%%%%%%%%%%%%%%%%%%%%%%%%%%%%%%%%%%%%%%%%%%%%%%%%%%%%%%%%%%%%%%%%%%%%%%%%%%%% %

There's another category of technology development that is more speculative but worth mentioning here. 
A number of companies like Brain Corporation, Evolved Machines, Grok --- formerly called Numenta, IBM and Vicarious have embarked on projects to build systems that make use of ideas from neuroscience to emulate brain-like computations. 
%
% http://www.braincorporation.com/, http://vicarious.com/, https://www.groksolutions.com/
%
Their business model is based on developing computational architectures patterned after the brain to solve practical problems in anomaly detection, visual perception, olfactory recognition, motor control and autonomous vehicle navigation. 

There is a long history of neuroscientists using existing simple electrical, mechanical and computational models to explain neural circuits~\cite{McCullochandPitts43,WalterSCIAM-50,StentetalSCIENCE-78,Braitenberg86}. It's not clear that we know enough about the basic principles governing real brains to engineer artificial brains.  Indeed it would seem that if those principles were widely known they would have been incorporated into state-of-the-art computer-vision and machine-learning systems given that those fields include leading experts in computational neuroscience, but that is manifestly not true~\cite{DiCarloandCoxTCS-07}.  There is another possible approach to unraveling the secrets of the brain that holds out more promise and that dovetails with the focus on this report on developing new scientific instruments to record neural activity at the scale of whole brains. 

% %%%%%%%%%%%%%%%%%%%%%%%%%%%%%%%%%%%%%%%%%%%%%%%%%%%%%%%%%%%%%%%%%%%%%%%%%%%%%%%%%%%%%%%%%%%%%%%%%%%%%%%%%%%% %

Even the near-term technologies that we have discussed in this report promise to provide unprecedented detail across significantly larger populations of neurons than possible previously.  Moreover with new tools from optogenetics and better optical methods for delivering light to deep tissues we now have the capability of selectively activating and silencing individual neurons in a given circuit.  Optogenetic tools are particularly convenient for probing neuronal sensitivity, mimicking synaptic connections, elucidating patterns of neural connectivity, and unraveling neural circuits in complex neural networks~\cite{CallawayandYusteNEUROBIOLOGY-02}.  It will soon be possible to instrument and record from neural circuits in awake, behaving subjects exposed to natural stimuli. The potential for hypothesis-driven science to discover the underlying principles is enormous, and the prospects for data-driven exploration may be at least as promising. 

Engineers routinely apply machine-learning algorithms to fit models with thousands or even millions of parameters to data.  Typically the data defines the inputs and outputs of a function that we would like to infer from the data and then apply to solve a problem such as analyzing an image --- the input is a numerical array of pixel values --- and determining if it contains an instance of a particular class of objects --- the output is true or false.  However, it is also possible that the function is realized as a physical system and the inputs and outputs correspond to measurable parameters of that system.  In this case, the objective is generally to infer some function that reproduces the behavior of the physical system or at least the observable inputs and outputs in the data. 

An investigator interested in inferring such a function might start with a hypothesis couched in terms of a family of models of some explanatory value, with success measured in terms of the fitted model accounting for observed data~\cite{KerretalSCIAM-08,ByrneetalNEURO-10}.  Alternatively, the investigator might be satisfied with a family of models expressive enough to capture the behavior of the target system but of little or no explanatory power and a set of experiments demonstrating that the fitted model accounts for the data, including held-out data not included in the training data.  This sort of relatively opaque model can be employed as a component in a more complicated model and the resulting compositional model might prove to be very interesting, both practically and scientifically. 

For example, a number of scientists believe that the neocortex is realized as a homogeneous sheet of anatomically-separated, computationally-similar units called {\it{cortical columns}}~\cite{MountcastleBRAIN-97,MountcastleCC-03}. If so, then if we were to infer the function of one cortical column in, say, the visual cortex, and then wire up a sheet of units implementing this function, then perhaps the composite sheet would exhibit the sort of behavior we observe in the early visual system.  This particular example is quite challenging, but the same principle could apply to a wide variety of neural circuits, e.g., the retina, believed to have this sort of compositional architecture~\cite{ZaghloulandBoahenVPOD-07,SaglametalNIP-08}. 

A team knowledgeable in large-scale machine learning might partner with one or more labs involved in recording from neural circuits and share data and expertise to essentially {\it{mine}} the data for algorithmic and architectural insight.  Products amounting to black boxes realizing useful adaptive strategies and pattern recognition behaviours would provide substantial value in a wired world in desperate need of such behaviors exhibiting the robust character of biological systems.  Scientists scour the world for novel genes and microorganisms that exhibit useful behavior and it is this trove of natural technologies that hold out such promise for the next generation of neural recording instruments~\cite{RuschetalPLoS-07}.  It may be that these same instruments will help to reveal an equally valuable algorithmic windfall for the companies with the wherewithal to harvest it.

% %%%%%%%%%%%%%%%%%%%%%%%%%%%%%%%%%%%%%%%%%%%%%%%%%%%%%%%%%%%%%%%%%%%%%%%%%%%%%%%%%%%%%%%%%%%%%%%%%%%%%%%%%%%% %

% Researchers in Johns Hopkins-led study find that portable eye-tracking device diagnoses stroke with 100 percent accuracy:

% http://www.hopkinsmedicine.org/news/media/releases/is_it_a_stroke_or_benign_dizziness_a_simple_bedside_test_can_tell

% Dr. Uzma Samadani, Shani Offen, David Heeger and Marisa Carrasco --- patent application for eye-trackng neurological-assessment tool:

% Shani Offen --- LinkedIn 
% Research Assistant Professor
% NYU Center for Neural Science
% January 2013 --- Present

% I developed a method for performing uncalibrated eye movement tracking that enables rapid and facile assessment of neurologically impaired patients. 

% Dr. Uzma Samadani and I, in conjunction with David Heeger and Marisa Carrasco, have submitted a provisional patent application for this technology. 

% ``To put Shani's contribution to the field in perspective, the Department of Defense has awarded several million dollars worth of funding since 2005 to at least two different laboratories to develop a mechanism for using eye movement tracking for neurologic assessment of brain injured patients.  Neither of these substantially larger labs working for more than half a decade has accomplished a fraction of what Shani has achieved in just over a year.'' --- US, Shani's manager at the VA.

%% file: acknowledging.tex
\section{Acknowledgements}
\label{section:acknowledging}

This report is the culmination of nine months of research into the topic of scalable neuroscience and, in particular, the prospects for developing new scientific instruments to accelerate research in the brain sciences.  During the first six months, a small band of us including Ed Boyden, Greg Corrado, David Heckerman, Jon Shlens, Akram Sadek and, toward the end of this period, Yael Maguire and Adam Marblestone collaborated to explore the space of relevant technologies and identify promising opportunities.  We also enlisted the help of a larger group of scientists many of whom participated in CS379C at Stanford and provided consulting to the students working on final projects.  In the final three months prior to releasing this document, the students in CS379C listed as coauthors on the report joined the effort identifying additional opportunities and analyzing those considered most promising.  We are particularly grateful to Ed Boyden, Kevin Briggman, David Cox, Mike Hawrylycz, Yael Maguire, Adam Marbelstone, Akram Sadek, Mark Schnitzer, Jon Shlens, Stephen Smith, Brian Wandell and Tony Zador for joining in our class discussions.  We would also like to thank Arjun Bansal, Tony Bell, Kwabena Boahen, Ed Callaway, John Donoghue, Bruno Madore, Bill Newsome, Bruno Olshausen, Clay Reid, Sebastian Seung, Terry Sejnowski, Dongjin Seo, Fritz Sommer and Larry Swanson for timely insights that served to adjust our perpectives and shape this report.

%% file: biafra.tex
\subsection*{Overview}

The BRAIN initiative, Brain Activity Mapping (BAM), Human Brain Project, and others like it seek to gather the activity and structure of many neurons; this data would help test comprehensive models of the brain and design better treatments for nervous system diseases, e.g. Huntington's, Dementia, Pain, and others. The human brain has on the order of $10^{11}$ neurons, ~$10^{14}$ synapses, over a hundred neurotransmitters, and dozens of genetically defined cell types among other parameters~\cite{azevedo2009equal}, which traditional technologies --- microscopy, electrophysiology, electrode arrays, and others --- are ill-equipped to deal with. This magnitude and specificity of data is normally overwhelming with without serious compromises: experiments are invasive and subject viability is reduced. In contrast, DNA sequencing is used to measure millions to billions of molecules, is readily scalable (both reading, storing, and analyzing the data~\cite{loh2012compressive}), and always getting cheaper~\cite{glenn2011field}. Due to this, next-generation sequencing technology is proposed to help connectome construction and activity recording. Doing this will require integration of several technologies, many of which are already available and ready to use while others still need some tweaking. Based on analysis of recent publications and currently available technologies, we are optimistic that sequencing can be used to produce connectomes and firing patterns in cell culture or simple model organisms in the near term (1-2 years) with use in rodent models in the medium term (3-5 years). However, a concern remains about the viability of this approach to study human connectomics without serious technical improvements, on both the science and policy side.

There are several proposals on how to use DNA sequencing to study connectomics and activity, we will focus on two and several technologies that would help in their implementation. To study the connectome, the plan is to individually barcode all neurons in a brain, allow the barcodes to spread to synaptically connected partners, ligate host and foreign barcodes, and sequence~\cite{ZadoretalPLoS-12}. To record activity, one proposal is to encode the activity as errors in a DNA template~\cite{KordingPLoS-11}. These are fundamentally molecular biology challenges that need to be overcome, however the possibility exists that using nanotechnology~\cite{AlivisatosetalNANO-13} will help improve the reliability and experimental breadth of these technologies by taking advantage of the increasing returns in computing power and size reduction as epitomized in Moore's~\cite{moore1965cramming} and Bell's law~\cite{bell1972effect} laws along with Intel's Tick-Tock model of chip architecture miniaturizationd~\cite{singhal2008inside}.
	
\subsection*{Technical}

Some of the technologies outline below are available (Recombinases, Super-resolution microscopy, FISH, etc.) while others still need to be fully developed (ion-sensitive and cytoplasmic DNA polymerases, induced DNA secretion, etc.). However, a timeline for when they can be integrated into a sequencing-based approach to BAM is of interest given the many parts needed to get the entire systems working.

\textbf{Ion-sensitive DNAPs} DNA polymerases (DNAPs) can be sensitive to ion concentration and a recent publication characterized a particular one, Dpo4, for its transfer function (relation between ion concentration and error rate) with $Mn^{++}$, $Mg^{++}$, and $Ca^{++}$. Dpo4 was only useful for discerning $Mn^{++}$ and $Mg^{++}$, which are physiologically less preferred than $Ca^{++}$~\cite{ZamftetalPLoS-12}. However, it is likely that ion-sensitive DNAPs can be found or created by mining microbiology literature and metagenomic searches\footnote{BLASTing for DNAPs that appear to have calcium binding domains or similar structural elements to DNAPs with known calcium sensitivity.} or through DNAP engineering (e.g. via directed evolution~\cite{ghadessy2001directed,xia2002directed,fa2004expanding,ghadessy2004generic}). For example, modifying DNAPs can yield increases in desired parameters, such as the addition of a T3 DNA polymerase thioredoxin binding domain to Taq polymerase that caused a 20-50 fold increase in processivity~\cite{davidson2003insertion}. Further, a basic search of Polbase, a DNA polymerase database, shows that T7, Pow, and Pab Pol I all have very low error rates (lower than Dpo4); in addition, T7 has high processivity and is quite rapid---13,404 bases/min or 223.4 Hz. Given that networks can fire between 40-200 Hz, this should allow enough resolution to pick up spiking at rest or low frequency activity~\cite{maex2003resonant}. The availability of many DNAPs and the pressure to improve them within the molecular biology community --- for use in PCR and other assays --- indicates that an ion-sensitive DNAP could be created or found within 1-2 years and implemented in mammals in 3-5 years.

\textbf{Cytoplasmic DNAPs} Transcription in the cytoplasm would be preferred for activity-based measurements using DNA polymerases to avoid the different characteristics of calcium transients in the nucleus~\cite{bootman2009update}. A special form of transcription that can occur in the cytoplasm has been observed and might be adapted for the ticker tape system.\cite{cheng2012plasma} Given the paucity of experimental data showing localization of DNAPs to the cell membrane --- either via anchoring to transmembrane proteins, addition of GPI anchors, or other methods --- it is unlikely that this component will be implemented until 3-5 years out unless a breakthrough in adapting cytoplasmic viral replication machinery, such as that of the Mimivirus, can be demonstrated.

\textbf{Modified DNAP trangenetic mouse} It takes at minimum around two years to make a trangenetic mouse line~\cite{conlon2011transgenic}. Given that ion-sensitive nor non-viral cytoplasmic DNAPs haven't been fully characterized and a cell line containing a barcode cassette has just been made\cite{zador2013personal} indicates that we are 3-5 years from obtaining a mouse model that natively expresses a modified DNAP, either constitutively or under control of specific inducible constructs (e.g. Cre, FLP, etc.).

\textbf{Recombinases} Cell lines with a \emph{Rci}-based shuffling cassette stably integrated to allow random barcoding of neurons has already been created~\cite{zador2013personal} and a randomized transfection library already demonstrated to work~\cite{oyibo2013probing}. Full characterization in cell culture and invertebrate model organism is likely 1-2 years out while use with rodents is 3-5 years out (see transgenetic mice). This technology is unlikely to be used with humans in the near future.

\textbf{2$^{\mbox{\rm{nd}}}$/3$^{\mbox{\rm{rd}}}$ generation sequencers} PacBio RS, 454 GS/FLX and Life Technologies Starlight\footnote{Not yet on the market.} have the longest read lengths needed for activity-based sequencing. However, Starlight isn't available yet, PacBio has an extraordinarily high error rate of around 12\%, and 454 GS/FLX series have position-dependent error rates, which are not preferred. While it is possible to implement the proposed BRAIN sequencing methods now, improvements in read length, error rate, position-dependent errors, and error type are needed to reduce the problem of biases. Because of the competitive environment and rapid pace of improvement in this area, we expect the needed technology is in the 1-2 year pipeline and if needed can be optimized specifically for these applications given adequate interest and funding.

\textbf{Super-resolution microscopy} STORM, PALM, STED, and other microscopy techniques have helped illuminate the fine structures of the cell and commercial systems are already available.\cite{SchermellehetalJCB-10} Integrating blinking fluorophores with GRASP, FISSEQ, and FISH for identification of synaptically connected neurons and \emph{in situ} sequencing is possible in the next 1-2 years and has been previously supported to occur given adequate interest.\cite{huang2010breaking}

\textbf{FISSEQ/FISH/ISH-HCR} These technologies have been improved over the past decade and incorporation into this system should be trivial~\cite{MitraAB-03,beliveau2012versatile}, with an estimate of 1-2 years. They each require slightly different approaches, but we envision FISSEQ or ISH-HCR winning out due to versetility in design of barcodes and ability to sequence error-strewn activity templates. Clear$^{\mbox{\rm{T}}}$ and CLARITY might provide a method of imaging without needing to arduously slice brains, but they are new techniques and at the moment not optimized. It will likely take 1-2 years given concerted effort and 3-5 years given parallel development to realize FISH or FISSEQ in 3D volumes without the need for slice work.

\textbf{GRASP} Another older technology that might be adapted for use in BRAIN sequencing~\cite{FeinbergetalNEURON-08}. The idea would be to synaptically couple two barcodes then sequence using FISH/FISSEQ and distinguish closely localized synapses using super-resolution microscopy. Skepticism remains about the ability to distinguish multiple points at the diffraction limit, but our calculations indicate that at an average of 1 synapse per cubic micron, this shouldn't be a large problem.

\textbf{Exosome-based DNA secretion} Cells secrete exosomes (small vesicles containing protein, DNA, and other small molecules) and they could be made to carry DNA containing connectome or activity data into the blood stream and out the renal system~\cite{bobrie2011exosome}. However, few mechanisms for experimentally inducing their secretion are known (calcium being one signal, but that would cause problems in a calcium detecting system). Technology to non-invasively alter nanoparticles exist for thermal, radio, and magnetic signals~\cite{HuangetalNATURE-10,stanley2012radio,bernstein2012optogenetics}, so it may be possible to couple these technologies to induce DNA secretion to allow the subject to be kept alive during readout. Due to the multiple systems that need to be put into place, this is optimistically a 3-5 year outlook.

\textbf{Engineered proteins} These would be proteins that respond to ultrasonic, thermal, radio, magnetic, and other non-invasive signals. This has already been demonstrated for ion channels, but awaits conjugation to DNAPs or other proteins for integration into BRAIN sequencing systems~\cite{HuangetalNATURE-10,stanley2012radio,bernstein2012optogenetics}. Given that pieces of the technology exists we estimate 3-5 years for proof-of-concept and 5-10 years for expression in rodents.

\textbf{Multiplexed probes} The possibility exists that multiplexing delivery of sequencing chemicals and readout of the signal is possible, seeing that it has already been done with neural recording and drug delivery~\cite{seidl2010plane}. For example, the drug delivery channel would uptake small samples of surrounding fluid and use a nanodevice with a DNAP molecular imprinted onto its surface. Binding and sequencing of local nucleotides could be offloaded to a nearby sequencer or a clever method could take advantage of changes in surface conductance upon binding\footnote{For example, how binding in surface plasmon resonance (SPR) changes reflectivity.} of specific segments of the DNA strand, measure the change, and read this out as a measure of the current nucleotide being sequenced (given some previously defined standard)~\cite{boyden2013personal}. Given that we can now either offload the heavy lifting to already available DNA sequencers and because the probes have already been fabricated, preliminary results could be seen in brain slices within 1-2 years.

\textbf{Nano-sequencers} Inserting sequencers \emph{in vivo} that don't need the addition of expensive reagents, high-cost library construction, and other add-ons associated with parallel sequencing\cite{glenn2011field} would yield benefits both from near real-time data acquisition and reduced complexity. Detecting proteins~\cite{cai2010molecular} and DNA~\cite{KorlachetalPNAS-08,MannionandCraigheadBIOPOLYMERS-07,CraigheadNATURE-06} using micro-technologies already exists and nanopore technologies should allow these to be scaled down~\cite{kasianowicz2008nanoscopic,ZhangetalSENSORS-13,heller2010next}. Already, theoretical calculations for DNA nano-sequencers has been demonstrated and the possibility of integrating this with other nano-scale read-out technologies (e.g. RFIDs or OPIDs) could allow sequencing without the DNA needing to leave the system~\cite{min2011fast}. Further, this would also improve the false-negative rate as DNA can be degraded or altered during transport out of the body. The fact that theoretical calculations for detecting base-pair differences sans pyrosequencing have been done but devices don't yet exist peg this as a 5-10 year technology.

\subsection*{Predictions}

The following table summarizes the best estimates of when key technologies discussed in this proposal will be implemented for use in DNA sequencing for connectomics or activity recording. These times represent optimistic estimates for proof-of-concept in cell-culture then onto animal models (add an extra 2-3 years). We do not consider use in humans as that is 10+ years away for sequencing-based technologies, partially because gene therapy also needs to advance to a point where the constructs designed for use in model organisms can be easily transfered with drastically diminished risk for neuronal death induced by over-expressing viral vectors and other complications. Because of this long timeline, human implementation is not directly relevant to BRAIN goals in the short/medium term for this set of technologies (in contrast to fMRI, EEG, and other medically approved, non-invasive technologies).
\begin{center}
\begin{minipage}{\textwidth}
\begin{center}
  \begin{tabular}{|c|c|c|c|}\hline
    Technology                         & Area        & Years  & Notes \\\hline
    Ion-sensitive DNAPs                & activity    & 1 to 2 & \footnote{DNA polymerase sensitive to Mn and Mg exist and are mentioned in Zamft~\etal~\cite{ZamftetalPLoS-12}.}\\\hline
    Cytoplasmic DNAPs                  & activity    & 3 to 5 & \\\hline
    Modified DNAP transgenetic mouse   & activity    & 3 to 5 & \\\hline
    Recombinases                       & connectome  & 1 to 2 & \\\hline
    2$^{\mbox{\rm{nd}}}$/3$^{\mbox{\rm{rd}}}$ generation sequencers & both  & 1 to 2 & \footnote{Need to see improvements in error rates and read length.}\\\hline
    STORM/PALM/STED                    & both        & 1 to 2 & \\\hline
    FISSEQ+2D slice                    & both        & 1 to 2 & \\\hline
    FISSEQ+CLARITY/Clear$^{\mbox{\rm{T}}}$ & both        & 3 to 5 & \footnote{The ability for 3D imaging to obtain diffraction limited sampling in unknown at present.}\\\hline
    FISH+2D slice                      & both        & 1 to 2 & \footnote{This would be more useful for connectomics where a known set of barcodes is possible. One could envision using the percent binding of a template probe to determine activity.}\\\hline
    FISH+CLARITY/Clear$^{\mbox{\rm{T}}}$   & both        & 1 to 2 & \\\hline
    GRASP                              & connectome  & 1 to 2 & \\\hline
    Exosome-based DNA secretion        & both        & 3 to 5 & \\\hline
    Multiplexed probes                 & both        & 3 to 5 & \\\hline
    Engineered proteins                & both        & 5 to 10 & \footnote{Ultrasonic microbubbles are currently being considered for gene delivery but few proteins are known to respond to it specifically.}\\\hline
    Nano-sequencers                    & both        & 5 to 10 & \footnote{Based on estimates from current lab-on-a-chip devices.}\\\hline
  \end{tabular}
\end{center}
\end{minipage}
\end{center}

%% file: mainak.tex
\subsection*{Overview}

We present some estimates of the scale at which we would be able to record from
neurons simultaneously. The treatment here is data modality agnostic, i.e. we
may have calcium imaging data, data from optical sensors, or data from direct
electrode measurements of neuronal voltages.  The focus here is more on the
potential of computer analytics to identify functional patterns with the
measured data.  The goal may be to identify functional collections of neurons
or achieve super resolution imaging based on the fact that the activities of
nearby neurons are correlated.

A central problem in deducing function/structure from measurements is the issue
of identifying which cell gives rise to a certain measured signal. Any measured
waveform is contaminated by waveforms from nearby cells. Spike sorting
algorithms try to rectify this by separating individual waveforms (from
different neurons) from their linear combination.  One can envision that these
spike sorting algorithms combined with sophisticated side-information will
enable us to record from more neurons than is possible now with electrodes. We
provide an example below to illustrate the kind of scaling possible and the
development time scales involved.

Of course such techniques would invariably be guided by the functional models
of the brain and the quality of measured data. In the following sections we
review some of the technologies and algorithms which show great promise in
helping us scale our readout problem. While most of the discussed technologies
are currently possible only in animal models because of the need for novel signal
measurement devices (requiring genetic enginering or highly invasive electrode
recordings), one can imagine that the signal processing algorithms used here
would be data agnostic and hopefully will carry over even to non invasively
gathered human brain data.

% Details including your analysis and conclusions: "How did you arrive at your
% conclusions?"
\subsection*{Technical}

% \subsubsection*{Recording techniques considered}

Most of the methods in this section focus on signal recording of neural
activity, either in the form of direct measurements of voltage levels, or
indirect measurement of the calcium concentration levels using genetically
encoded calcium indicators (GECIs). Although the latter usually suffer from lack of
resolution (mainly temporal) some GECIs like GCaMP3 and/or GCaMP5~\cite{AkerboometalNEUROSCIENCE-12} offer much faster kinetics (i.e. greater
temporal resolution) and greater stability across interesting timescales (i.e.
greater readout duration). As mentioned later, such properties are crucial for
increasing the readout duration from neurons, and may serve as convenient
replacements for micro electrode arrays for high throughput brain signal
measurement.  These signals are used in conjunction with  sophisticated microscopy (laser or fluorescence)
and/or sophisticated algorithms/models to perform parameter estimation. The
hope is that prior knowledge of the structures or processes generating the
measured signals would either help in the refinement of the estimated
parameters or in scaling up the number of signals read. In fact if the sizes of
the neurons are large enough (e.g. some hippocampal and cerebellar cells) and
if the functional regions are stable enough (across behavioural states and time
scales as is the case with hippocampal place cells), such techniques offer
practical ways to scale up the number of neurons one can image. 

% \subsubsection*{Reconstructing from measured signals} 

One of the main issues in making sense of measured signals from the brain is
identifying which cells they come from. While rate of spiking was long
considered to be critical for understanding neural function~\cite{stein2005neuronal} , it was slowly realized that other waveform
properties (like relative spike timing patterns) may also be important in
encoding neuron function. This is where spike sorting algorithms come in. These
try to separate linear superposition of signals into waveforms for each neuron.
Algorithms like PCA (principal component analysis), ICA (independent component analysis), and particle filter (sequential Monte Carlo methods)
have been successfully used in some settings to recover spiking patterns of
individual cells. While these algorithms are not free from artifacts, these
techniques in combination with other modalities like cell-body segmentation do
offer powerful ways of inferring connectivity patterns and/or functional
classification from signals~\cite{JainetalCON-10}.

% \subsubsection*{Some quantitative estimates of current and future capabilities} 

\subsection*{Predictions}

In general the exact reconstruction techniques used depends on the area of the
brain we are imaging and the signals or parts of the brain we are interested
in. Techniques like the ICA (independent component analysis) have been used to
good effect in understanding biomedical signals (EEG, MEG, fMRI)~\cite{vigario2000independent}. Of course, such analysis tend to be based on
some assumptions about the underlying signal generation process, in particular
about measured waveforms being a linear combination of several independently
generated waveforms. While this in isolation may not correspond to realistic
models of neural function, it does offer useful insights especially when
combined with different techniques~\cite{jung2001imaging}.  The number of
sample points needed however scales as the square of the number of independent
components we are trying to recover. One can think of applying a similar
concept in the analysis of neural signals. The independent components would
correspond to different functional units. Combining measurements from different
approaches may help to annotate reconstructed signal with e.g. location
information from microscopy. 

Using ICA, we generally need of the order of $n^2$ data points to be sampled in order to
estimate the mixing coefficients of the $n$ independently firing neurons. Thus
for approximately $80-100$ neurons in a cortical minicolumn or $7-50$ neurons
in a cerebellar microzone (an anatomical and functional collection of neurons),
we need to be able to record on the order of $10^4$ data points. Since neural
connectivity patterns or functional evolution can be significantly affected by
timeshifts of a few milliseconds~\cite{markram1997regulation}, we need at least
one millisecond temporal resolution. This entails that we record at least for
$10 s$ or more. While this is not hard with direct electrode recordings (which
often have good temporal resolution), the use of indirect (e.g.  calcium
imaging) techniques often need sophisticated signal processing to achieve the
same (e.g. to deduce precise spike times from calcium levels).  Thus not only
do we need to record from many neurons simultaneously, we also need to be able
to record for a longer duration per neuron. Due to plasticity of synaptic
connections and functional associations and the photobleaching effects (or instability) of the indicators, this may be a challenge in some
settings. However, in some cases mentioned below where there is stability of
the functional units across time or behavioural states, the problem is solvable
using the approaches mentioned here.  

We note in passing that while these ideas are representative of the gains that
can be achieved, they are not the only ones considered in the literature. The
problem of spike timing inference has been looked at from other perspectives
also. For example,~\cite{VogelsteinetalBJ-09} uses particle filtering to deduce
spike times. With simulated data they conclude that using models can outperform
standard estimation methods like Wiener filters. However, for brevity and to
convey representative estimates of how much we can scale, we henceforth focus on
specific work.

We consider the work  in \cite{mukamel2009automated}, and look into the
evolution of similar techniques.  By using techniques like independent
component analysis together with cell body segmentation, the authors in~\cite{mukamel2009automated} could recover signals from $> \mathbf{100}$
Purkinje cells.  By analysing the correlation patterns, they could also
identify microzones with sharp delineations (width of one Purkinje cell). This
level of precision was not achieved by direct measurement methods at the time
it was published due to the coarse (approximately 250$\mu$m) inter-electrode
spacing.  Future work in~\cite{dombeck2010functional} used a combination of two
photon imaging and its correlation with virtual reality patterns to achieve
functional imaging of hippocampal place cells at cellular resolution during
virtual navigation.  By using a different microscopy technique, researchers in~\cite{GhoshetalNATURE-11} were able to obtain simultaneous recordings from
$> \mathbf{200}$ Purkinje cells simultaneously across different microzones.
This was scaled to $>\mathbf{1000}$ hippocampal pyramidal cells in~\cite{ziv2013long}. Of
course, the absolute numbers are dependent on which part of the brain we image
and the specific techniques we use, but the numbers do reveal the power of more
refined models of neural activity and improved signal recording capabilities in understanding and
better characterising the functional state of the brain.

In the remaining part of the discussion we present some estimates for the
potential of better algorithms in helping us understand other functional units
of the brain which may not necessarily be as stable as the microzonal
structures in the cerebellum. A topic of intense research in this is the
cortical minicolumn~\cite{BuxhoevedenandCasanovaBRAIN-02}.  Proposed in~\cite{mountcastle1957modality} in 1957 this has led to significant research in
identifying groups of neurons as a functional unit, instead of focussing on
single neurons. It talks about functional organization of cortical neurons in
vertical columns, the idea being that they would be activated by the
stimulation of the same class of peripheral receptors.  This idea is the basis
for much of future work. For example, the Blue Brain Project~\cite{MarkramNATURE-06} aims to simulate the human cortical column on the Blue
Gene supercomputer. The basic workhorse is still the Hodgkin Huxley model,
simulated on a supercomputer to reveal interesting macrostructures.  While
there is strong evidence for cortical microcolumns being important units of neural
function and pathways, they do show some sort of plasticity, i.e. the neurons
from different cortical columns may organize themselves into units which change
with time and functional stimulus~\cite{BuxhoevedenandCasanovaBRAIN-02}. Of course
having the model is not enough. One needs to be able to fit the free parameters
in the model to the empirically measured data. This is one of the main focus
areas for the Blue Brain project also. Readouts from patch clamp techniques or
other MEA (microelectrode array) recordings are used to model the different
neurons.  Other techniques like spectral clustering (which is similar to PCA)
offer some insights~\cite{eldawlatly2009identifying}.

In humans, these minicolumns range in length from $28 \mu m$ to $40 \mu m$.
While electrodes offer ways to record from neurons in different parts of the
brain simultaneously, they usually suffer from poor spatial resolution (but good
temporal resolution). However, some of the latest developments in electrode
technologies offer a solution to that. Using the resolution offered by the
latest 3D optogenetic arrays, we can get close to 150 $\mu m$~\cite{ZorzosetalOPTICS-12}, but other
techniques like fluorescence microscopy~\cite{GhoshetalNATURE-11} offer
feasible ways of imaging cortical columns in a level of detail which would be
enough to resolve neurons to single cell precision. 

In short these techniques in conjunction with microscopy and other data
acquisition techniques offer scalable and non destructive  ways of increasing
spatiotemporal resolution (as compared to using them in isolation e.g.
microelectrodes which have good temporal resolution but poor spatial resolution
or  traditional imaging tools which generally have poor temporal
resolution-although some recent work~\cite{GreweetalNATURE-10} promises sub
millisecond accuracy). In any case, these techniques can help not only to
reconstruct from measured data, but also to guide data acquisition.
Incorporation of available side-information about the neurons we are interested
in can offer powerful ways of scaling up the number of neurons we are able to
image.

%% file: anjali.tex
\subsection*{Overview} 

Nuclear magnetic resonance (NMR) imaging modalities are currently the most promising technologies for recording macroscale measurements, with resolution on the order of millimeters and seconds, used in the analysis of the major functional areas and the white-matter pathways connecting those areas in awake, behaving human subjects. Magnetic resonance imaging (MRI) sensitive to quantitative NMR tissue properties, diffusion, and blood oxygenation are currently the tools of choice for studies of normal and pathological behavior in the field of cognitive neuroscience as well as clinical diagnosis.

Diffusion tensor imaging (DTI) and quantitative MRI (qMRI) have already been used in \emph{in vivo} human studies to quantify fascicle and tissue development, and therefore cognition and behavior. DTI models the distribution of diffusion directions of water protons as a tensor, providing measures of the apparent diffusion coefficient (ADC) and fractional anisotropy (FA). Identical lobes of a gradient in the diffusion-sensitive direction are applied separated by a 180 degree pulse and a temporal wait so that diffusing spins, unlike stationary spins, are not fully re-phased and thus contribute attenuated signal. Quantitative mapping of the proton density (PD), which is proportional to the amount of water, and the $T_{1}$, which measures the spin-lattice relaxation constant and thus quantifies the interactions between protons and their molecular environment, can be used to measure Macromolecular Tissue Volume (MTV), the non-water volume in the voxel, and the Surface Interaction Rate (SIR), the efficiency of a material's energy exchange with water protons~\cite{MezerMTV}.  Since water preferentially diffuse along axons, DTI enables tractography of the fascicles and since approximately 50\% of the macromolecules in the white matter are myelin sheaths, MTV measurement provides an indication of axonal diameter and myelination. Although there is no dynamic activity mapping, the changes in structural properties give an indication of function due to long-time-scale development and plasticity of the brain. Specifically, the rate of change of FA, MTV, and SIR in the posterior corpus callosum, the arcuate fasciculus, and the inferior longitudinal fasciculus have been shown to correlate with reading ability~\cite{WandellandYeatmanCOiN-13}.  In addition, changes may be used to diagnose and evaluate diseases; for example de-myelinating disorders such as multiple sclerosis can be identified using MTV measures.  

Functional MRI (fMRI) enables the coarse spatial and temporal localization of neural activity through the proxy of the hemodynamic blood oxygen-level dependence (BOLD) response. After neural activity, ions must be pumped across cell membranes for repolarization. This necessitates blood flow to the region, resulting in elevated oxygenation levels in the 2 to 3 millimeters surrounding active neurons 2 to 6 seconds after they fire. Oxy-hemoglobin is diamagnetic like most substances in the body, while deoxy-hemoglobin is paramagnetic. Therefore, the increase in oxy-hemoglobin levels results in a decrease in the magnetic susceptibility differences between the blood and the surrounding tissue. Since the variation in the proton resonant frequencies correspondingly decreases, $T_{2}*$ lengthens, and the image appears more intense in the active regions. Taking advantage of this indirect coupling of a magnetic spin parameter to synaptic activity, it is possible to map the brain activity in response to motor and cognitive tasks, thereby enabling the better understanding of these functions at the macroscopic level. Although the seconds-scale response time is much slower than many neural processing dynamics, the information from fMRI has still proven very valuable. For example, fMRI has facilitated the identification of the visual field maps in the human visual cortex and improved the understanding of the perception and function of the visual system~\cite{WandelletalNEURON-07}. 

The advantages of MRI stem from its noninvasiveness, its endogenous contrast, and its scalability, and therefore the fact that it can be used in human studies. Although MRI does not have neuron-level resolution, it does have the ability to measure more macroscopic variables that reflect properties of the local population of neurons and how these local populations interact. It thus has the potential to address the “missing length scales” in neuroscientific knowledge --- while neurons and synapses are reasonably understood, we do not have a good understanding of the microarchitectural units in the brain and especially how these units communicate, interact, and work together in a global network. Therefore, since we do not yet have the ability to understand the complexity of individual neuron data across the entire brain, courser spatial-scale information is valuable for increasing the understanding of the brain and is scalable in terms of both acquisition time and data processing and storage. MRI in particular can be used to study the integration of signals across brain circuits and the development of the white matter, and therefore cognition, over time and experience. Especially in these areas of research, recording from humans is necessary, as smaller, less-developed animals lack the extensive pathways and complex cognitive functions of humans~\cite{WandellPC}.  Therefore, the promise of MRI primarily stems from its ability to be used on humans to study the global brain network. 

\subsection*{Technical}

Currently, scan times of 8 to 12 minutes were necessary to achieve 1.5 mm$^{3}$ to 2 mm$^{3}$ resolution at 1.5 T and \textasciitilde{}1 mm$^{3}$ resolution at 3.0 T for qMRI and DTI~\cite{MezerMTV}.  The long duration of MR image acquisition posed a challenge in prior studies, preventing the study of cognitive development before the age of eight~\cite{WandellPC}.  Although fast sequences have been used for both qMRI and DTI, the acceleration factors from parallel imaging, if any acceleration was used at all, were relatively conservative~\cite{MezerMTV}.  Therefore, there is still a lot of room for improvement in the use of parallel MRI (pMRI) systems to facilitate the greater understanding of the development of the brain at the macroscale. A system with four receiver coil elements and a corresponding four will be utilized in the very near future to scan six to eight year olds~\cite{WandellPC}.   

The fMRI signal, stemming from susceptibility effects of blood oxygenation, is typically very weak. Therefore, signal averaging is necessary to achieve satisfactory image quality, and high main field strengths are utilized to increase the signal strength, and therefore the signal-to-noise ratio. Acquiring the signal multiple times to facilitate averaging, however, increases the scan time by a factor of the number of averaged signals. The acceleration from parallel imaging, therefore, can keep the image acquisition duration reasonable, especially for pediatric imaging. Especially at the high field strengths used for fMRI, the acceleration factors obtained from pMRI are high. In addition, parallel imaging ameliorates the tradeoffs incurred from increasing the main field strength.$ T_{2}*$ generally shortens as field strength increases, causing the signal to dephase quickly during readout~\cite{pMRI}. Since the BOLD response is rapid, however, fast sequences with extended readout times such as echo planar imaging (EPI) are used for fMRI. Therefore, the shortened readouts provided by parallel coils are particularly valuable in fMRI to avoid the decay of the signal. Parallel imaging has already been successfully used for functional imaging~\cite{de2006accelerated}.  Gains from parallel imaging have not been exhausted, however, and improvements in pMRI over the next 2-3 years are predicted to significantly contribute to functional neuroimaging.

Coil arrays with up to 32 elements which can achieve up to an eight-fold improvement in scan time without significant degradation of the image quality. Using current RF coil hardware, if the acceleration factor is increased further, the sensitivities of the coil elements used for each measurement overlap spatially and the measurements are no longer necessarily independent in the presence of noise. It thus becomes difficult to solve the inverse problem of reconstructing the image from the set of acquired signals and coil sensitivities as the effective rank of the relevant matrix (resulting from the signals and sensitivities) decreases. In addition, the number of elements in RF coil arrays are limited by a signal-to-noise ratio (SNR) penalty. Since smaller coils are more sensitive to the surface than to deeper structures, reconstructing the image at the center of the brain requires summing the small signals from all of the array elements. In doing this, noise is also received from all of the coils (rather than primarily from the nearest surface coil, as is the case for superficial structures), and the SNR degrades to become comparable to that of a body coil~\cite{NishimuraPC}.  As better RF coils are designed, improving the spatial selectivity and uniformity (over the limited volume) of the elements, coil arrays with 256 elements and an acceleration factor of 32 can be expected within the next 2-3 years. This improvement may enable the testing of subjects as young as 2 or 3, which would, for example, allow the study of cognitive development that begins earlier than reading, as well as the development of motor skills, which occurs at too young of an age to be studied with current technologies. In addition, in other applications where fast scanning is not essential, the scan time improvement can be traded in for SNR and resolution improvements. 

Improvements in image acquisition time and signal to noise ratio, as well as the application of the most advanced current technology to neuroscience studies, therefore, will very probably contribute to the understanding of the brain's structural development and function within the next three years.  

\subsection*{Ultrasound}

The lack of an acoustic window into the brain is a significant roadblock to the use of ultrasound or noninvasive neural imaging in humans.  A back-of-the-envelope illustration of this uses the characteristic impedance mismatches between air, bone, and tissue and therefore the reflection coefficients, the frequency necessary for desired resolution, and the attenuation coefficient at this frequency (using the~\textasciitilde{}1 dB/MHz/cm loss rule of thumb). These determine the intensity necessary for the signal to be above the electronic noise floor at the penetration depth into the brain desired. The intensity at the focus, combined with the absorption coefficient of the tissue, gives the specific absorption rate (SAR), and the bioheat equation can be used to determine the increase in temperature from the SAR. Using the Arrhenius damage integral, the temperature function can be converted into an equivalent thermal dose at 43 degrees C, which gives an indication of if the tissue has been damaged via coagulative necrosis. In addition, the intensity of the pressure waves can be compared to thermal and mechanical limit indices to evaluate the safety. Due to the acoustic properties of the skull, ultrasound may be more suitable for neural stimulation using low-intensity, low-frequency ultrasound, or therapeutic ablation using high intensity focused ultrasound (e.g. of the thalamus for essential tremor treatment) rather than imaging.  The main contribution of imaging to this field, therefore, may be MR, which provides thermometry measurements and good contrast, to guide and monitor the use of focused ultrasound.  An alternative may be to surgically insert transducers, but, in doing so, some of the main advantages of imaging, noninvasiveness and therefore the ability to conduct human studies, are lost.  Since the main challenges to using ultrasound in neuroscience is not temporal or spatial resolution, but depth penetration/SNR and heating issues, despite recent, near-term, and intermediate-term improvements for rapidly scanning tissue at high temporal (dynamic) and spatial resolution, it is likely not most promising technology for brain imaging.

%% file: andre.tex
\subsection*{Overview}

We discuss in the following section the use of micron-scale, implantable optical devices for wireless readout. We present an optical identification tag (OPID), an analog of the conventional Radio-frequency identification tag (RFID), as a sensor for neuronal monitoring of the firing of axon potentials as well as a communication means with other devices.  After a brief description of the structure of OPIDs, including their necessary size and their components,  we present two technological schemes for wireless readout via nanotechnology, each leveraging the use of OPIDs and potentially realizable in the next 4-8 years. In each, we envision a region of the brain (potentially the entire brain) where an OPID is placed next to each neuron such that it can record, in real time, the firing of the axon potential. 

The first readout system would utilize larger, yet still micron-scale, implanted RFID chip reporters to relay information from the sensor OPIDs to devices external to the brain. Each reporter would communicate locally with the sensors nearest to it, and would then transmit an aggregated signal, containing the data of its local sensors, to an external receiver, where the data could be processed. This system would allow for real-time monitoring of the firing of large regions of neurons. The OPIDs could be installed via the methods described in the introduction, and the larger RFID reporters would be surgically implanted. The second readout system is similar to the first, but uses an insertable fiber-probe to communicate with the sensors. Each fiber probe would contain hundreds, or even thousands of chips capable of communicating with the sensors, and would remain connected to some external circuitry. However, instead of relaying the information wirelessly, each of these reporter chips would send data along the probe itself via electronic circuits.

\subsection*{Technical }

With Moore's Law holding over the past many decades, transistors have reached dimension sizes of 22 nm and are projected to decrease to 10 nm by the year 2016. As such, the design of an RFID with micron dimensions has been achieved to well within the dimensions of neuronal cells~\cite{burketowards2010}, given their size range of 4 to 100 $\mu m$. In fact, the latest in RFIDs can fit inside the larger ones. Assuming a linear scaling of RFID size with transistor size, we can expect a 10,000 transistor RFID to decrease to about 10 $\mu$m $\times$ 10 $\mu$m $\times$ 5 $\mu$m by the year 2016.

One type of RFID, proposed by Yael Maguire at Harvard, is an optical frequency RFID, or OPID, measuring 10 $\mu$m $\times$ 10 $\mu$m $\times$ 5 $\mu$m and containing~\textasciitilde{}10,000 transistors that would sit either inside a neuron or immediately adjacent to it. Each OPID will contain a neuronal voltage sensor~\cite{peterkaimaging2011}, sensitive enough to detect voltage levels on the order of millivolts, with a time-resolution within 1 millisecond. The OPIDs will operate while under the constant illumination of the reporters and utilize small optical components capable of communication with either larger RFID chips or fiber-probes. When the OPIDs sense axonal firing, a CMOS circuitry layer with a CPU will store this information, and at specified intervals dynamically change the load impedance of their optical components to modulate backscattered signals. Maguire suggests the powering of such a device using a high-efficiency solar cell. Wireless power transfer via magnetic induction and glucose powering are two other avenues for powering these devices~\cite{kurswireless2007}. State of the art glucose energy extraction has achieved 1.0-3.3 $\mu$W/cm$^2$ or 1.0-3.3 $\times$ $10^{-2}$ pW/$\mu$ m$^2$~\cite{hansenhybrid2010}. Transistors today operate at picowatt scales, so either the glucose extraction efficiency or the transistor operating power need to increase/decrease by about 3 orders of magnitude. Extrapolating Moore's law, we can expect glucose powering to become feasible in 16 years.

\textbf{RFIDs as Reporters:} We now consider the first system, in which OPID sensors communicate with local RFID reporters that send information outside the brain. These OPIDs will modulate the backscatter of light sent by the RFIDs. We choose optical communication in the wavelength range of 600 nm to 1400 nm, the {\it{biological window}}, due to its low absorption coefficient~\cite{rogganoptical1999} of 0.1-0.5 mm$^{-1}$ and its high frequency, enabling low-loss communication with substantial bandwidth. This choice results in a tradeoff with scattering. Light scattering at these wavelengths is quite strong, and leads to a very small coherent penetration depth. Grey matter in the brain suffers from a scattering coefficient of approximately 10 mm$^{-1}$~\cite{yaroslavskyoptical2002}, corresponding to a signal-to-noise ratio (SNR) of $e^{-10}$ for two objects separated by 500 $\mu$m. White matter has an even greater scattering coefficient of approximately 30 mm$^{-1}$, corresponding to a SNR of $e^{-30}$ at the same separation distance. Given this constraint, an OPID in grey (white) matter will need to communicate with a separate reporter within a distance of about 500 $\mu$m (150 $\mu$m).

As RFIDs decrease in size, RFID-based implantable biomedical devices continue to decrease as well. Pivonka~\etal~\cite{pivonkaa2012} report a 2 mm $\times$ 2 mm wirelessly powered implant with both a communication channel in the low GHz frequency range and a magnetic induction powering range of 5 cm. The device itself spans only 600 $\mu$m $\times$ 1 mm $\times$ 65 nm, with the larger dimensions being the result of the coil used for wireless power transfer. A mere 500 $\mu$W is all that is needed to continuously power the device, and this takes into account the power that goes into the locomotive motion of Pivonaka's design - a stationary device would require substantially less power. This powering distance range, along with this power threshold, allows for access to the majority of the brain. In fact, the powering distance is the main limiting factor in how deep into the brain an RFID can be placed. Communication distances are not an issue provided the chosen frequency band has low extinction in brain tissue, and provided that the RFID is large enough to contain a dipole or folded dipole antenna functional in this frequency band~\cite{fana2007}. Therefore, it is reasonable to expect, in the next 4-8 years, a decrease in the size of the magnetic induction coil relative to the wavelength to the point where the same range of 5cm could power a coil of 500 $\mu$m $\times$ 500 $\mu$m, given the decrease that has occurred since wireless power transfer was derived~\cite{kurswireless2007,pivonkaa2012}.

Having addressed the issues of powering RFIDs and OPIDs, as well as defining frequency windows of communication for each, we next turn to the interaction between the sensor OPIDs and their reporter RFID counterparts: We assume the implantation of a 10 $\mu$m $\times$ 10 $\mu$m $\times$ 5 $\mu$m OPID for every neuron in some region of interest (ROI) of the human brain, yielding a sensor density of $8 \times 10^4$ sensors / mm$^3$. This leads to a $4\%$ increase in the volume of the ROI, or identically a $4\%$ decrease in extracellular fluid. For illustrative purposes we consider a region of grey matter, where a sensor-reporter pair can be distanced up to 500 $\mu$m apart. Assuming the insertion of RFIDs on the order of 500 $\mu$m $\times$ 500 $\mu$m $\times$ 30 $\mu$m spaced every 1 mm, then each reporter would be linked to all sensors within a 1 mm$^3$ rectangular volume surrounding it. Given that the firing rate of a neuron is approximately 10 Hz, then we can expect 800 spikes/ms on average. 

A given OPID will need to transmit its own identity in the form of some serial number, timing information about the firing of the axon potentials, and an additional amount of overhead accounting for error correcting codes, redundancy, etc. If the OPID transmits a packet of data every time its neuron fires, it will need to transmit about 100 bits per packet, assuming 12 bits to contain firing information, 37 for an OPID identification number, and just over 50 for overhead. For a 1 mm$^3$ ROI, this corresponds to $8 \times 10^5$ total signals sent per second, (sps) and $8 \times 10^7$ bits/second (bps). While the bps is small, the sps could lead to congestion at the reporter end, similar to bad cell phone reception in a densely crowded area. Suppose that, instead of an OPID transmitting every time its neuron fires, it stores the data in a buffer and transmits it every second, yielding a sps of $8 \times 10^4$ sps. The necessary buffer could be achieved with Maguire's 10,000 transistors, in 4 years in accordance with Moore's law. Reducing the sps by another factor of ten to $8 \times 10^3$ sps would require 10 times the buffer size. This is accomplishable, for the same-sized device, in 8 years time.

Next we must consider the issue of bandwidth between the RFIDs and OPIDs. As mentioned above, the biological window provides an 800 nm wavelength ($\lambda$) range in which transmission can occur. As an illustrative example, we consider frequency-division multiplexing between the OPIDs. That is, each OPID is assigned, by the RFID, a specific frequency band in which it will communicate, in order to distinguish between signals. In communication systems, the $Q$-factor $Q=f/\Delta f$ is the ratio of the frequency of transmission to the half-max bandwidth, and is a measure of frequency selectivity. For 800 OPIDs around 1 RFID each utilizing 1nm $\lambda$ of bandwidth, each OPID will need to transmit with a photonic device offering $Q=1.05 \times 10^3$. For $4 \times 10^4$ OPIDs per RFID in a volume of 1 mm$^3$, photonic devices with $Q=52.5 \times 10^3$ will be needed. While this may be achievable with quantum dots in years to come, this is possible to achieve at the moment with very thin photonic crystal slabs. Specialized photonic crystal cavities have demonstrated ranges from $Q = 45,000$~\cite{akahanehigh-q2003} to $Q \geq 1,000,000$ in the wavelengths of interest~\cite{asanoanalysis2006}. Another solution, also proposed by Maguire, is the use of an LCD screen in the OPID that can modulate the backscatter of an impinging electromagnetic (EM) wave. It has been shown that by applying a small voltage, the reflectance and efficiency of an LCD display can be boosted to $\geq 95\%$~\cite{komandurireflective2008}. 

\textbf{Electronic Fiber Probes as Reporters:} Using RFIDs as reporters introduces the additional challenge of wireless relay from the RFIDs to external devices. RFIDs will work well for small ROIs, but the bit rate transmitted from them scales linearly with volume. In the limiting case of the entire brain and each OPID transmitting at the above $8 \times 10^7$ bps, the total data rate will be $2 \times 10^{14}$ bps, or 2 Terabits/sec. Instead of attempting to handle an  enormous amount of data wirelessly, it may be more feasible to replace the RFIDs with thin electronic fiber probes, on the order of 500 $\mu$m in diameter. Suppose a ROI in the brain had OPIDs attached to every neuron, and further suppose that a series of probes, in parallel, were surgically implanted into this ROI. By spacing the probes with the same characteristic spacing of the RFIDs (\textasciitilde{}1 mm$^3$), and placing RFID-like chips inside the probes, at this same characteristic spacing, it would be possible to read data from the neurons and transmit information electronically up the probe instead of dealing with it wirelessly. The drawback to this system is the added volume needed in which to place the entire probe.

%% file: daniel.tex
\subsection*{DNA Sequencing Implants}

Nanotechnology is poised to offer much benefit in sequencing DNA, and this benefit can be harnessed for brain activity mapping.  There are many proposals for using DNA to record information from neurons, such as recording synaptic spikes and mapping the connectome --- see Appendix~\ref{appendix:biafra}.  We expect other information (such as topography) also will eventually be recordable in DNA.  Once information is recorded in DNA, however, getting the information out still presents a formidable challenge.  One possibility is to encapsulate the DNA in a vesicle, which would migrate through the extracellular fluid to DNA sequencing chips inside the brain, at which point the DNA would be sequenced and the information would be read out digitally.

The migration of the vesicles from neurons to the chips creates a surprising amount of difficulty.  These vesicles would probably have diffusivities similar to other such vesicles, or around the order of 10$^{-8}$ cm$^{2}$/s, causing them to be rather slow~\cite{Sheets2008Biophys,Hinton93}.   If 1 million sequencing chips were evenly distributed throughout the brain, it would take around a day for these vesicles to travel the distance needed to reach the chips.  A better solution (one that would both decrease the time and number of chips) would be to equip these vesicles with molecular motors.  Fitting these vesicles with molecular motors such as flagella could give them effective diffusivities of around 10$^{-5}$ cm$^{2}$/s~\cite{Bush2010}.   With only 1000 chips evenly distributed throughout the brain, it would take on average less than 2 hours for these augmented vesicles to traverse the necessary distance.  1000 chips is a small enough amount that they could be implanted manually, though we expect such a procedure to be automated.  The vesicles additionally would contain functional groups to allow them to target to these chips.  A large number of each of these vesicles would need to be released for each corresponding strand of DNA to ensure that at least one makes it to a chip.  Once the vesicle makes it to the chip, it would release the DNA for sequencing.  

Sequencing the DNA could be performed quickly by nanopore technology. Nanopores consist of small holes in a membrane (such as in silicon or graphene).  As DNA is threaded through this nanopore, voltage readings on the membrane indicate translocation events that correspond to specific base pairs, allowing for fast sequencing.  Currently, each pore can read 300 bp/s (base pair per second)~\cite{DNAGuardian}.   A chip containing an array of $100 \times 100$ nanopores reads 3 Mbp/s.  Assuming DNA sequencing follows its exponential increase in speed (doubling time of about 0.8 years), this leaves us with speeds of around 200 Mbp/s and 12 Gbp/s for these chips in 5 and 10 years~\cite{Futrealnature}.   Assuming 1000 chips in the brain and estimating that a bit of data might take tens of bp to encode, then in 5 years this system would be able to hand 10 Gbit/s, and in 10 years it could handle 1 Tbit/s.  The 5 year option presents 1 bit for every 10 neurons per second for readout, which is likely too small for mapping activity for anything other than short experiments, but could be used to determine other information, such as cell types or connectomics.  By 10 years from now, this technology would have 10 bits per neuron per second.  In theory, this would allow for mapping the activity of 10\% of the neurons in the brain with 10 ms resolution, or 1\% of the neurons in the brain with 1 ms resolution.  If trends continue, all 100B neurons could be readout with 1 ms resolution in about 15 years. Finally, this information would be read out from the chips using fiber optics.  With each chip processing 10 Mbit/s in 5 years and 1 Gbit/s in 10 years, this is clearly within the range that fiber optic wires could handle.

\subsection*{F\"{o}rster Resonance Energy Transfer}

F\"{o}rster Resonance Energy Transfer (FRET) is a phenomenon that can be exploited to optically map the activity of the brain.  FRET can occur when two chromophores are close to each other; the donor chromophore transfers its energy to the acceptor chromophore by dipole-dipole coupling, leading the acceptor to fluoresce.  If the chromophores are not close, the donor fluoresces at a different frequency.  Optical imaging techniques can therefore be used to determine if the chromophores are within a certain distance.  Placing these chromophores either on different molecules or specific locations of the same molecule allow one to determine if the two molecules are interacting or if the one molecule has undergone a conformational change, and this can be harnessed for brain activity mapping.  The two most promising approaches for using FRET are a genetic engineering based approach and a nanotechnology based approach.

Genetic engineering has already led to some successes in using FRET for mapping brain activity.  For instance, Cameleon is a genetically encoded calcium indicator (GECI) that in the presence of calcium ions undergoes a conformational change which increases the FRET effect~\cite{pmid9278050}.   Cameleon can therefore be used to visualize synaptic spikes.  GCaMP5s (GECIs similar to Cameleon) are possibly the most advanced of such genetically encoded indicators and have been used to image the firing of over 80\% of the neurons in an entire zebrafish brain {\it{in vivo}} at 0.8 Hz and single-cell resolution~\cite{AhrensetalNATURE-13}.   With each iteration of the GCaMP molecule, there have been a few large improvements of properties (such as three-fold increases in contrast from GCaMP3 in 2009 to GCaMP5G in 2012) and a number of smaller improvements~\cite{AkerboometalNEUROSCIENCE-12} --- see Appendix~\ref{appendix:nobie}.  Extrapolation of these trends indicates that these molecules will undergo incremental improvements in coming years.  The genetic engineering approach to FRET has the advantages of being more mature than the nanotechnology approach and being capable of implementation in a relatively easy, nontoxic manner.  However, the requirement for genetic engineering does present a potential barrier for use in humans.

The nanotechnology approach to FRET typically involves quantum dots (QDs), nanoparticles that confine excitons in all three spatial dimensions and thus exhibit only a few allowed energy states.  QDs can be used in FRET as either the acceptor or the donor (or both), and have been used to image action potentials~\cite{NadeauetalEMBS-06}. QDs have many beneficial properties for use in FRET.  For instance, QDs have broad absorption spectra with narrow emission spectra, and varying the size of QDs substantially varies these spectra.  Therefore, they are bright, tunable, and have high signal to noise~\cite{ZhangandWangTHERANOSTICS-12}.   The biggest detriment against using QDs for FRET is probably their toxicity.  There are currently efforts to reduce toxicity by coating QD in polymers or functionalizing them with ligands, but these efforts so far have only seen partial success~\cite{Nienature2004}.   Another hurdle will be introducing the QDs into the brain.  Focused ultrasound can be used to temporarily disrupt the blood-brain barrier, allowing for the QDs to cross into the brain~\cite{WangandHuSCIENCE-12}.

The genetic engineering based approaches to FRET seem likely to dominate for at least the next five years, due to their relative maturity and problems QDs face.  Nanotechnology ultimately has more to offer, and probably will overtake the bioengineering approaches after toxicity and delivery hurdles are solved, likely between 5 and 10 years from now.  For both of these approaches, imaging in mammals and in particular people presents issues largely due to the opaqueness of the brain.  Microendoscopy or related technology could be used to overcome these problems --- see Appendix~\ref{appendix:oleg}.

\subsection*{Carbon Nanotube Neural Stimulation}

In addition to mapping brain activity, nanotechnology has the potential to control the activity of the brain with great precision.  The realization of such capabilities has many market and social incentives.  In addition to treatment or cures for neurological disorders, fine control promises benefits in many endeavors, including gaming, learning, and augmentation.  Furthermore, controlling the activity of the brain will be instrumental in mapping the activity of the brain, as control will allow us to discern causality instead of simply correlation.  Carbon nanotubes (CNTs) are one class of molecule that can be exploited for this control.

One such scheme involves using DNA to interface CNTs with ion channels.  CNTs can be separated by length using centrifugation~\cite{Fagan2008}.   These different length CNTs can then be wrapped in specific DNA sequences that target particular chiralities of CNTs, such that each sequence corresponds to CNTs of a particular length and chirality~\cite{pmid19587767}.   These resultant complexes (DNA-CNT) are nontoxic and able to cross the blood brain barrier~\cite{pmid20652100}.   These DNA-CNT would be fitted with strands of DNA on their ends such that the strands target particular ion channels.  This targeting will probably initially require the ion channels to be genetically engineered so that the DNA-CNT have an easier time homing in on them, but we envision that eventually the DNA-CNT could be functionalized to specifically bind to non-engineered ion channels.  By differing the targeting DNA strands depending on the length of the CNT, different length DNA-CNT will target to different types of ion channels~\cite{SadekNotes}.   CNTs oscillate when irradiated with light (with oscillation depending on length of the CNT), and this oscillation in turn activates the ion channel~\cite{SadekNotes}.  By using different frequencies of microwave light, we can target different length CNTs and thus different ion channels~\cite{SadekNotes}. Using only semiconducting CNTs and not metallic CNTs --- which are dependent on the chirality --- prevents the CNTs from heating up under this irradiation~\cite{SadekNotes}.

This process would have a lot of the same capabilities as optogenetics, but would also have certain advantages.  For one, this method would allow for many more types of ion channels to be independently activated.  This is because the narrow range of light usable for optogenetics would allow for only a handful of ion channels to be independently activated without much crosstalk.  The carbon nanotubes, on the other hand, could be fabricated to different lengths such that a very large number of channels could be independently activated~\cite{SadekNotes}.  Additionally, microwaves can penetrate the skull, while visible light cannot, so this method can be used without implants.   Furthermore, we might be able to target DNA-CNT to ion channels without genetic engineering, thus circumventing a major obstacle for use in people.  Of additional note, DNA-CNT is slowly broken down by the body, so this technique would not be permanent.

%% file: nobie.tex
\subsection*{Overview}

The technologies presented in this section all fall into the category of near-term opportunities with potential for incremental progress over the longer term. Non-invasive means for detecting neural structure and activity is paramount for both basic neuroscience and clinical therapies. Novel contrast agents for MRI and photoacoustic tomography (PAT) present technologies that can be developed to reliably gain large scale (whole-brain) structure and activity information. A holy grail for all imaging techniques is to enable identification of specific molecules. Several avenues for using contrast agents to identify molecules are being explored in these techniques. All of these technologies are available for use in animal models. Potential for clinical use is noted where applicable. 

Genetic engineering has matured to the level of complexity development. DNA can be deterministically created and inserted into a genome. There already exist genetically encoded indicators for a variety of neurotransmitters. Finally, optogenetics provides a tool to optically excite or silence neurons. All of these pieces have brought genetic engineering to a point of complexity development. The leading question here is, how can one combine these techniques to learn more about the biology of the brain? As these techniques have already been developed, incremental progress will proceed with a tilt toward complexity of implementation. Finally an important class of genetically encoded indicators  are calcium indicators, also known as genetically encoded calcium indicators (GECIs). Currently GECI's present important information regarding neuronal activity, but do not present activity on the level of a single action potential. The current state of the art GECI, Gcamp5, exhibits great improvement over previous versions and incremental advancement is expected. 

Novel tissue preparation techniques like CLARITY~\cite{ChungetalNATURE-13} and Clear$^{\mbox{\rm{T}}}$~\cite{KuwajimaetalDEVELOPMENT-13} present new ways for researchers to optically probe prepared tissue. Previously, researchers needed to cut small portions of the brain, stain, and image this small area. With CLARITY and Clear$^{\mbox{\rm{T}}}$, one can stain large portions--an entire mouse brain --- entire brain and image separate areas confocally. These techniques also allow for washing and restaining of the same tissue. Incremental progress will proceed in refining the technique, and research groups will ramp up use of the tool.

\subsection*{Technical}

Focused Ultrasound (FUS) has been combined with photoacoustic tomography for deep (3mm in brain tissue) and accurate imaging of blood brain barrier (BBB) disruption~\cite{WangetalJBO-12}. In this work, gold nanorods (AuNR) were used as a contrast agent for PAT to image the time dynamics of BBB disruption by FUS. Gold nanorods are particularly well suited as a contrast agent for photoacoustic tomography (PAI) due to their tunable optical properties and the potential for gene delivery~\cite{Wu2006}. Additionally, gold nanocages have been explored as contrast agents with capability for drug delivery~\cite{Wang2009}. 

Photoacoustic Tomography produces images by using ultrasound and the transduction of absorbed photons into heat and subsequently pressure. It has been used in mice for functional imaging of hemodynamics without the use of a contrast agent. It has the advantage over MRI of imaging both oxygenated and de-oxygenated hemoglobin, two values correlated with cancer and brain function~\cite{Wang2003}. Additionally it can be combined with contrast agents for molecular targeting. It is compatible currently available molecular dyes~\cite{Wang2008IEEE} that have potential for approval for clinical use. Finally, PAT has potential for clinical use as it is non-invasive and can image intrinsic absorbers.  To date, PAT has achieved imaging on many scales --- from organelles to organs --- with the capability of imaging up to 7 cm into brain tissue with sub-millimeter resolution~\cite{WangandHuSCIENCE-12}. PAT can be implemented in a variety of different manners; it has been used to image blood flow (as in doppler photoacoustic tomography)~\cite{Wang2008MedPhys}, gene expression~\cite{Wang2007BiomedOptics}. Many implementations of PAT are intrinsic and noninvasive. Thus, they present opportunities for human studies and clinical use. PAT has already been commercialized for preclinical use through VisualSonics and Endra.  It is predicted that PAT usage will increase within 1-2 years for preclinical applications and rapid adoption and commercialization for clinical use will follow upon FDA approval. 

Magnetic Resonance Imaging provides another method for non-invasive measurement of neural activity. Bioengineering of contrast agents serves a powerful method to develop contrast agents; one could imagine building a versatile indicator for MRI that is similar to the green fluorescent protein (the workhorse of molecule targeted optical microscopy). Magnetic nanoparticles such as super paramagnetic iron oxide when conjugated with calmodulin can be use as a calcium indicator for neural activity. Directed evolution has been used to develop a dopamine indicator based on a heme protein~\cite{Jasanoff2010}.  Finally, enzyme based contrast agents combining MRI contrast with in situ chemical processing. An early example visualized \(\beta\)-galactosidase expression via enzymatic hydrolysis of a Gadolinium substrate~\cite{Meade2000}. Given the mature level of this technology incremental development of molecular targeted contrast agents will occur. 

GCaMP, a popular GECI, has become an indispensable tool for neuroscience
research. The state of the art, GCaMP5, cannot reliably detect sparse
activity--one test setup provided detection of single action potentials (APs)
26\% of the time~\cite{AkerboometalNEUROSCIENCE-12}. Previous versions of GCaMP had negligible
detection of single APs. Dynamic range has also rapidly increased GCaMP2,
GCaMP3 and GCaMP5H exhibited \(\Delta F/ F =\) \(5.1\pm 0.1\), \(12.3\pm 0.4\),
and \(158 \pm 12\) respectively. Finally, calcium affinity has also increased
over previous versions of GCaMP. The rapid progress of the GCaMP family of
indicators and the recent release of GCaMP6 (on AddGene) indicates that this
technology will continue to develop the potential for calcium imaging of sparse
activity. Furthermore, GCaMP has been used in mice, drosophilia, zebrafish, and
C. elegans and can be presented as a viral vector or can be expressed
transgenically. GCaMP is already commonly used in neuroscience laboratories to
image neural activity. However, many advancements remain. Thus we place this in
the short term advancement and implementation category with the hope that
reliable detection of single action potentials arrives within 3-5 years.         

The recent introduction of CLARITY and Clear$^{\mbox{\rm{T}}}$ to the neuroscience community presents a great tool and will quickly revolutionize experimental neuroscience~\cite{ChungetalNATURE-13}. The technique allows for multiple rounds of in-situ hybridization and immunohistochemistry. Combining CLARITY with optogenetics and/or calcium imaging can provide detailed information regarding activity, structure, and expression. Clear$^{\mbox{\rm{T}}}$ presents a powerful tool for developmental biology and interrogating. At the moment it is unclear in what ways the technique will be used. However, the latest tissue clarification techniques open new possibilities for research and will become commonplace in laboratories within 1-2 yrs. 

%% file: oleg.tex
\subsection*{Overview}

In this section we talk about implantable devices for activity recording in dense neural populations. The focus here will be particularly on recordings from deep brain regions as opposed to superficial layers that are easier accessible by other means. We first give the estimates for the scalability of microendoscopy approach. We then investigate avenues for development of next generation of implantable devices that can further boost the number of recorded locations and individual neuron cells per location.

\textbf{Microendoscopy:} Optical ways to probe and stimulate neural activity are gradually taking over traditional methods such as multi-electrode arrays. This is facilitated by great advances in GECI and more recently voltage sensitive fluorescent proteins.  A number of fluorescence microscopy techniques utilizing these advantages have been employed to record neural activity. Among them: single- and two-photon excitation scanning microscopy~\cite{denk1997photon}, light-sheet microscopy~\cite{AhrensetalNATURE-13} and others. Due to severe light scattering, imaging depth is limited to $\sim$500-750 um, so only superficial brain tissue is reachable for conventional microscopy techniques~\cite{wilt2009advances}. The inside of the brain machinery is, of course, as essential as the periphery and the main challenge is to obtain high-quality recordings from deep brain regions as well.

It has been shown that implantation of a carefully engineered microendoscope allows extending capabilities of standard microscopy techniques to regions arbitrarily deep in the animal brain. Microendoscope  composed of several GRIN (Gradient Index) lenses relays the high quality optical image above the animal scull, which can be further accessed by conventional optics. Strikingly, activity from more than a thousand cells in a mouse Hippocampus can be recorded simultaneously using this technique~\cite{ziv2013long}. Both single and two-photon imaging modalities are possible using GRIN endoscopes~\cite{KimetalJBO-08}. Thus, scaling the microendoscopy approach will be extremely useful to allow recording from multiple dense populations of neurons in several locations along the neural pathways in the brain. This should greatly facilitate studies of connectivity between different brain areas as other techniques do not provide this kind of resolution and coverage. Since most technological problems have been extensively debugged in single-endoscope experiments, it is reasonable to expect steps in direction of multiplication of microendoscopy recordings in the same animal. We believe that first attempts will be made in 1-2 year perspective and further increase in the number of simultaneous endoscope implantations will follow in 2-5 years. However, one of the main disadvantages of microendoscopy is its inherent invasiveness. This also becomes a limiting factor for recordings from too many endoscopes, since brain function can be substantially altered by implant. In the following technical section, we estimate the total number of simultaneously implanted endoscopes for deep-structure imaging not to exceed 20, which corresponds to recordings from $\sim$20000 neurons simultaneously. Though this number is just an order of magnitude larger than what is currently available with other technologies, it still opens unprecedented experimental opportunities, thanks to dense recordings at distant locations.

\textbf{Fiber-coupled microdevice implants:} Since scaling of microendoscopy recordings has its limitations, mainly due to high invasiveness, different approaches and technologies will emerge to keep up with the demand to obtain neural activity recordings at cellular level from more and more brain locations and more cells per location as well. In a general way we here discuss the advantages of using an optical fiber as an implant that can solve two major technological problems. First, sufficient power has to be supplied to the 'recorder` of neural activity. In case of ion or voltage sensitive florescent proteins excitation light has to illuminate neurons; in case of electrode arrays one needs to power all the electronics. In both cases optical power can be sent directly into the fiber. Second, optical fiber may have an extremely large bandwidth and serve to transmit optical signal encoding neural activity out from the brain.

Fiber implantation has been used in a variety of optogenetic experiments~\cite{BernsteinandBoydenTCS-04}.
However, stimulation with a single fiber lacks spatial resolution. Some attempts to construct fiber bundles have been made~\cite{hayashi2012spatio} which is though as bulky but inferior in quality to microendoscopy. Recording fluorescent signal  with a fiber itself is also not feasible as it does not carry spatial information, so optogenetics field is moving into direction of microfabricated multiwaveguide arrays~\cite{BernsteinandBoydenTCS-04,ZorzosetalOPTICS-12}.
We suppose that fiber implantation may become extremely valuable if coupled to a miniature device located on its tip. We envision the following functions such microdevice needs to possess:
\begin{enumerate}
\item microdevice needs to be well interfaced with the fiber to form a single minimally invasive implantable unit,
\item monitoring by optical or electrical means activity from several hundred to several thousand of surrounding neurons,
\item interface with the fiber, namely encode the spatial information and neuron firing times/patterns into an optical signal that is transmitted back into fiber, and 
\item efficiently manage optical power sent into the fiber from outside: direct incoming optical power to specific spatial locations (in case of fluorescence excitation) or convert it into an electrical signal to power built-in electronics.
\end{enumerate}

In the following subsection, we propose several implementation options and performance estimates. In the 5-10 year range, this approach may allow recording a comparable to microendoscopy number of cells per locations while being much less invasive.

\subsection*{Technical}

\textbf{Microendoscopy:} Typical microendoscope for imaging deep brain structure includes a micro-objective and a relay lens is $\sim$5mm long and $\sim$0.5 mm in diameter. This results in damaging of minimum 0.2\% of total brain mass of a mice (typically 0.4-0.5g). This accounts only for the volume that has to be replaced by an implant. In reality, damage might be significantly larger due to immune response to a foreign body. A number of studies examined different parameters of electrode implants that affect the damage to the brain.  Such factors as implant material and size as well as the insertion speed determine the number of damaged neuron cells~\cite{polikov2005response,thelin2011implant,biran2007brain}. Many of the findings can be extrapolated on microendoscope implants, however, to the best of our knowledge, no comprehensive studies have been conducted to quantify the effect of multi-endoscope implants from both microscopic and behavioral points of view. Though in some experimental protocols damage to certain areas of brain tissue is tolerable, one should not aim at more than 20 simultaneous deep-imaging sites, which will result in brain lesion of $\sim$4\% by volume.

By extrapolating density of recordings from~\cite{ziv2013long} we come to a conclusion that engineering advances can bring the microendoscopy to simultaneous recording from $\sim$20000 neurons deep in mouse brain. However, further increase of the number of endoscopes will potentially result in unreasonable complication of experiment and different deep-imaging approaches are needed. A system capable of handling and manipulation of such big number of microendoscopes in conjunction with simultaneous imaging would require significant engineering effort especially in miniaturization of optomechanics, though technology itself is available. We believe that due to extremely high interest and investment in the field, these issues may be solved in the nearest perspective, and functioning multi-endoscope imaging systems appear available to the neuroscience community in 1-2 years, and more advanced systems in 2-5 year perspective.

\textbf{Fiber-coupled microdevice implants:} The idea of a microdevice recording neural activity and coupled to an optical fiber is very generic and can be possibly implemented in a variety of ways and employing different technologies. For example, the sensing part of the device may consist of multi-electrode arrays or interface with different nanoscale recorders scattered in the brain (see above for details). Another option is to record activity information in an optical way directly, e.g., building a microdevice with a sensor that preserves spatial information: a combination of micro-optics and CMOS sensors, for example. Ideally, the lateral size of the microdevice should be matched to the diameter of the fiber, while constraints on its length are not as strict. This should be done to ensure that the whole implant causes minimal damage to the brain. 

According to~\cite{thelin2011implant}, implants 50 $\mu$m in diameter lead to larger survival fractions of neurons compared to bigger implants. To be more specific, we assume that the fiber with cladding and a device have a diameter of 100 $\mu$m. This would result in 25-100 times less tissue damage than in case of typical microendoscope of the same length.  A 50 $\mu$m or 62.5 $\mu$m diameter multimode fiber has a bandwidth of 10Gb/s, which is well above the bandwidth needed even for a very high resolution data one can be recording in a single location, e.g., $1000 \times 1000$ pixels image, 8-bit depth and 100Hz update frequency results in 0.8Gb/s.

An emerging field of Photonic Crystals (PhC) may greatly facilitate all-optical signal routing and processing on the microdevice itself. Efficient PhC-based waveguides, splitters and couplers and all-optical switches~\cite{joannopoulos2011photonic} have lateral footprint of $(\lambda/n)^2$, so a thousand distinct waveguides (one per recorded neuron) can easily fit laterally in the microdevice each addressing a distinct 'recorder` or a portion of the sensor.
In terms of power consumption the fiber can transmit up to several watts of optical power, so the limitation would be set by the amount of power that brain can dissipate (typically several tens of mW per imaging location). PhC all-optical switches also have an advantage over electronic counterparts as they can work in sub-femtojoule per bit regime~\cite{nozaki2010sub} (to power 100000 devices at 10 MHz one needs only 1 mW of power). The whole implanted device can be rapidly switched between input/output modes: fraction of the time for power input and the rest for recorded data streaming.
Of course, the exact architecture of a microdevice needs to be carefully engineered, but achieving similar to microendoscopy numbers of recorded cells should be feasible within 5-10 years, while significantly reducing the invasiveness.

%% file: ysis.tex
\subsection*{Overview}

We are interested in the applications of machine learning and robotics to automate tasks previously carried out by scientists and volunteers. This brings speed and consistency to experiments, but this also comes with questions and demands in error rates and efficiency. This is especially important in the brain readout problem, where an error rate of less than 0.01 percent can result in drastically skewed data once brought to a larger scale, eg. from a neuron to the connectome. In evaluating emerging technologies, we looked for improvements in scalability.  The technologies chosen to focus on all minimize the risk of scalable error. Furthermore, recognizing scalability as both vertical and horizontally applicable, we weighed technical value on more than strides in the error rate.  Considering horizontal scalability to be the technology's potential to expand into a human model and the primary obstacle to this being the invasiveness of the technology itself, we chose to focus on three technologies ranging in progression toward a noninvasive technique: patch-clamp electrophysiology, automated scanning electron microscopy, and high-throughput animal-behavior experiments. 

{\bf Scanning Electron Microscopy (SEM):}
Scanning electron microscopy generates three-dimensional images using a combination of two-dimensional images generated by focusing an electron beam across the surface of a biological tissue sample and collecting data on the backscattered electrons. This technology has seen tremendous advances in the last few years and is at the forefront of today's imaging methods. Furthermore, as electron microscopy staining has shown to be successfully unbiased in the staining of membranes and synapses in a neuron, in principle, the technology has the potential to be quite successful in mapping neural networks. There are three technologies within SEM that are automated, and as their automation accuracy and resolution improve, they are all viable techniques in reconstructing neuronal connectivity. These are serial block-face SEM (SBEM), automated serial-section tape-collection scanning electron microscopy (ATUM-SEM), and focused ion beam SEM (FIB-SEM). These methods range in imaging resolutions and dimensions, but all have the ability to be automated in a way that is scalable to reconstruct dense neural circuits. Once the challenge of increasing spatial scope and resolution is addressed, the main challenges that remain for SEM in automation itself are increasing the imaging speed.   It seems likely that in the next one to two years, acquisition will be fast enough to image an entire human brain in a reasonable amount of time (under one year). 

{\bf Patch-Clamp Electrophysiology and Probe Insertion:}
Patch-clamp electrophysiology is a robotic tool to analyze the molecular and electric properties of single cells in the living mammalian brain. Automation of the patch clamp technique began in the late 1990s and current patch-clamp algorithms allow for the high throughput detection, electrical recording, and molecular harvesting of neurons.  Recent advances, including a robotic process that allows for patch-clamp electrophysiology in-vivo, coupled with the ability to obtain information about the position or type of cellular structures being recorded indicate that this is a viable candidate for mapping neuronal connectivity. The major challenges to patch-clamp electrophysiology are throughput and the success rate of whole-cell recording. Nonetheless, there are significant breakthroughs currently being made in the field and these drawbacks will likely be resolved in the next five years.

{\bf High-Throughput Animal-Behavior Experiments:}
High-throughput animal-behavior experiments offer a means of studying human systems through an animal model. A challenge in using an animal model is acquiring and analyzing enough data to demonstrate that an animal model is an adequate comparison; therefore high throughput data collection using robotics is a viable way to expedite progress in this direction. Though animal-behavior experiments are less often considered at the forefront of viable technologies in mapping neuronal connectivity, development of new animal models exploiting characteristics of novel organisms may bring great advances in their parallels to the circuitry of certain parts of the human brain. The challenge of high-throughput animal-behavior experimentstion lies in being able to draw direct parallels from the animal to human models in brain circuitry, especially beyond the proven models in the visual system. In the next one to two years, expect to see an increase in high-throughput animal-behavior experiments as they are proving to be a viable option. 

\subsection*{Technical}

{\bf Scanning Electron Microscopy (SEM):}~\textit{Increasing Imaging and Acquisition Speed:} Highly parallel SEM is a development whose implementation is highly feasible in the next one to two years. It is possible to parallel imaging across multiple microscopes by assigning each to separate imaging sections. This would increase imaging speed by over two orders of magnitude. Another feature commonly overlooked is the overhead amount of time it takes to section, load, and unload specimens. This can result in a per-section overhead of up to six minutes, which in the case of ATUM-SEM is over ten percent of the per-section imaging time. Automating parts of this overhead component is achievable within the next one to two years and would increase imaging speed up to ten percent. In addition, a recent study in SBEM used a novel technique for coating specimens in scanning to eliminate distortion in electric fields due to the accumulation of negative charge; the technology is applicable to other automated SEM technologies~\cite{coating}. The overall result of such is the possibility to saturate camera sensors in a single frame, thus increasing system throughput. Coupled with the technological advances in image frame readout inherent in MooreÕs law, camera acquisition speed may nearly double in the next one to two years~\cite{briggman}.

{\bf Patch-Clamp Electrophysiology and Probe Insertion:}~\textit{Increasing Throughput and Whole-Cell Patch Recording Success Rate:} Current automated patch-clamp electrophysiology technologies can detect cells with 90\% accuracy and establish a connection with such detected cells about 40\% of the time, which takes 3-7 min in vivo. It is also worth noting that the manual comparison was a success rate of about 29\%~\cite{KodandaramaiahetalNATURE-12}. Nevertheless, the low successful connection rate limits the amount of data being collected and in order to achieve more comprehensive recordings, the area may require a great deal more sampling. In addition, the 3-7 minutes of robot operation is over a small localized area: scaling this method up to the entire brain may prove to require a lot more time, especially in the overhead of setting the robot up. This would be an undesirable amount of time for an in vivo study to take place. There are advances in increasing the speed of patch-clamp electrophysiology. In order to increase throughput, the use of multi-electrode arrays and multiple pipettes is being explored. This would increase throughput by as much as two orders of magnitude. In addition, groups such as Bhargava~\etal\ have worked to increase the success rate of obtaining whole-cell patch recordings through the use of a smart patch-clamping technique which combines patch clamp electrophysiology and scanning ion conductance microscopy to scan the cell surface and generate a topographic image before recording. This allows for microdomains and, consequently, a spatial functional map of surface ion channels~\cite{smart-patch}. They have also looked into expanded the size of the probe after surface mapping to increase the likelihood of capturing channels at those locations; they found a substantially greater yield in functional data on membrane features by increasing patch pipette size~\cite{super-res}. This has yielded a success rate of about 70\%, a dramatic increase from 40\%~\cite{high-res}. If this method can be incorporated in vivo, patch clamp electrophysiology will become a viable method for reconstructing neural networks.  

{\bf High-Throughput Animal-Behavior Experiments:} In the ideal situation, high-throughput animal behavior experiments may be used to model complete human neural networks. This would allow for extended observation and the option to work with more invasive technology. There is currently very little work being done in using these experiments for a complete model, but there has been extensive work in using animal models for the visual system. Mark Schnitzer's massively parallel two-photon imaging of the fruit fly allowed for as many as one hundred flies to be recorded at one time~\cite{briggman, serial-imaging}. Advances are also being made in using mice to model the human visual system; there is evidence of invariant object recognition in mice~\cite{ZoccolanetalPNAS-09}, as well as multifeatural shape processing~\cite{shape-processing} and transformation-tolerant object recognition.   The bottleneck is not the extent to which we can put animal experiments in parallel; it is in establishing parallels to investigate. At this pace, there will be significant advances in mapping the visual system in the next two to five years using animal experimentation. A complete neural circuitry, however, may not be accomplished using this method for the next five to ten years.